\documentclass[twocolumn]{aa}
\usepackage{graphicx}
\usepackage{setspace}
\usepackage{natbib}
\bibpunct{(}{)}{;}{a}{}{,}
\begin{document}
\setlength{\voffset}{0.5in}
\title{Atomic data for the K-vacancy states of Fe~{\sc xxiv}}
\author{
M.A. Bautista\inst{1}
\and
C. Mendoza\inst{2}\fnmsep\thanks{\emph{Present address:}
                                 Centro de F\'{\i}sica, IVIC, Caracas 1020A}
\and
T.R. Kallman\inst{2}
\and
P. Palmeri\inst{2}\fnmsep\thanks{Research Associate, Department of Astronomy,
                          University of Maryland, College Park, MD 20742}
}

\offprints{T.R. Kallman, \email{timothy.r.kallman@nasa.gov}}

\institute{Centro de F\'{\i}sica, Instituto Venezolano de Investigaciones
Cient\'{\i}ficas (IVIC), PO Box 21827, Caracas 1020A, Venezuela
\and
NASA Goddard Space Flight Center, Code 662, Greenbelt, MD 20771, USA}

\date{Received ; Accepted }


\abstract{ As part of a project to compute improved atomic data
for the spectral modeling of iron K lines, we report extensive
calculations and comparisons of atomic data for K-vacancy states
in Fe~{\sc xxiv}. The data sets include: (i) energy levels, line
wavelengths, radiative and Auger rates; (ii) inner-shell electron
impact excitation rates and (iii) fine structure inner-shell
photoionization cross sections. The calculations of energy levels
and radiative and Auger rates have involved a detailed study of
orbital representations, core relaxation, configuration
interaction, relativistic corrections, cancellation effects and
semi-empirical corrections. It is shown that a formal treatment
of the Breit interaction is essential to render the important
magnetic correlations that take part in the decay pathways of
this ion. As a result, the accuracy of the present $A$-values is
firmly ranked at better than 10\% while that of the Auger rates
at only 15\%. The calculations of collisional excitation and
photoionization cross sections take into account the effects of
radiation and spectator Auger dampings. In the former, these
effects cause significant attenuation of resonances leading to a
good agreement with a simpler method where resonances are
excluded. In the latter, resonances converging to the K threshold
display symmetric profiles of constant width that causes edge
smearing. \keywords{atomic data -- atomic processes -- line
formation -- X-rays: spectroscopy} }

\titlerunning{Atomic data for Fe~{\sc xxiv}}
\authorrunning{M.A. Bautista et al.}

\maketitle


\section{Introduction}
The iron K lines are among the most interesting features in astronomical
X-ray spectra. These lines appear in emission in almost all natural X-ray
sources, they are located in a relatively unconfused spectral region and have a
well-known plasma diagnostics potential. They were first reported in the
rocket observations of the supernova remnant Cas A \citep{ser73},
in X-ray binaries \citep{san75,bec77}, and in clusters of galaxies \citep{ser77},
the latter thus manifesting the presence of extragalactic nuclear processed
material. Observations of the galactic black-hole candidate Cyg X-1 showed
that the line strength varied according to the spectral state
\citep{bar85,mar93}, and \citet{tan95} found that the Fe~K lines
from Seyfert galaxies were relativistically broadened and redshifted which
suggested their formation within a few gravitational radii of a black hole.

Recent improvements in the spectral capabilities and sensitivity
of satellite-borne X-ray telescopes ({\em Chandra}, {\em XMM--Newton})
have promoted the role of Fe~K lines in
diagnostics, a trend that will continue to grow with the
launch of future instruments such as {\em Astro-E2} and {\em Constellation-X}.
Such plasma diagnostics ultimately rely on the knowledge of the microphysics of
line formation and hence on the accuracy of the atomic data.  In spite of the
line identifications by \citet{see86} in solar flare spectra and the laboratory
measurements of \citet{bei89,bei93}, \citet{dec93} and \citet{dec95,dec97},
the K-vacancy level structures of Fe ions remain incomplete as can be concluded
from the
critical compilation of \citet{shi00}. With regards to the radiative and Auger
rates, the highly ionized members of the isonuclear sequence, namely
Fe~{\sc xxv}--Fe~{\sc xxi}, have received much attention, and the
comparisons by \citet{chen86} and \citet{kat97} have brought about
some degree of
data assurance. For Fe ions with electron occupancies greater than 9,
\citet{jac80} and \citet{jac86} have carried out central field calculations
on the structure and widths of various inner-shell transitions, but these
have not been subject to independent checks and do not meet current
requirements of level-to-level data.

The spectral modeling of K lines also requires accurate knowledge of
inner-shell electron impact excitation rates and, in the case of photoionized
plasmas, of partial photoionization cross sections leaving the ion in
photoexcited K-vacancy states. In this respect, \citet{pal02} have
shown that the K-threshold resonance behavior is dominated by
radiation and Auger dampings which induce a smeared edge. Spectator Auger
decay, the main contributor of the K-resonance width, has been completely
ignored in most previous close-coupling calculations of high-energy
continuum processes in Fe ions
\citep{ber97,donnelly00,ber01,ball01}.  The exception is
the recent $R$-matrix computation
of electron excitation rates of Li-like systems
by \citet{whi02} where it is demonstrated that Auger damping is important
for low-temperature effective collision strengths.

The present report is the first in a project to systematically compute improved
atomic data sets for the modeling of the Fe K spectra. The emphasis is both on
accuracy and completeness. For this purpose
we make use of several state-of-the-art atomic physics codes to deliver
for the Fe isonuclear sequence:
energy levels; wavelengths, radiative and Auger rates, electron
impact excitation
and photoionization cross sections.
Particular attention is given
to the process of assigning reliable accuracy rankings to the data sets
produced.
Specifically, in the present report we have approached the radiative
and Auger decay manifold of the $n=2$ K-vacancy states of
Fe~{\sc xxiv}
as a test case of the numerical methods and the relevance of the
different physical effects. By detailed comparisons with previous work,
it has become evident that there is room for improvement, and that
an efficient strategy can be prescribed for the treatment of the whole
Fe sequence.
Furthermore, we also compute inner-shell electron impact excitation rates of
Fe~{\sc xxiv}, the total photoionization cross sections of
Fe~{\sc xxiii} and the partial components of the latter
into the K-vacancy levels of Fe~{\sc xxiv} where the relevant effects of
radiative and Auger dampings are fully established.


\section{Breit--Pauli Hamiltonian}
We have found the Li-like Fe system to be an unusually versatile
workbench for the magnetic interactions, a fact that perhaps has
not been fully appreciated in previous work. Thus prior to the
description of the numerical details of the codes, we include a
concise summary of the relativistic Breit--Pauli Hamiltonian
which is used throughout our computational portfolio and will be
central in the discussion of results.

The Breit--Pauli Hamiltonian for an $N$-electron system is given
by
\begin{equation}
H_{\rm bp}=H_{\rm nr}+H_{\rm 1b}+H_{\rm 2b}
\end{equation}
where $H_{\rm nr}$ is the usual non-relativistic Hamiltonian. The one-body
relativistic operators
\begin{equation}\label{1bt}
H_{\rm 1b}=\sum_{n=1}^N f_n({\rm mass})+f_n({\rm d})+f_n({\rm so})
\end{equation}
represent the spin--orbit interaction, $f_n({\rm so})$, and the non-fine
structure mass-variation, $f_n({\rm mass})$, and one-body Darwin,
$f_n({\rm d})$, corrections. The two-body corrections
\begin{eqnarray}\label{2bt}
\lefteqn{H_{\rm 2b}=\sum_{n>m} g_{nm}({\rm so})+g_{nm}({\rm ss})+}\nonumber \\
       & & +g_{nm}({\rm css})+g_{nm}({\rm d})+g_{nm}({\rm oo})\ ,
\end{eqnarray}
usually referred to as the Breit interaction, include, on the one hand, the
fine structure
terms $g_{nm}({\rm so})$ (spin--other-orbit and mutual spin--orbit) and
$g_{nm}({\rm ss})$ (spin--spin), and on the other, the non-fine
structure terms: $g_{nm}({\rm css})$ (spin--spin contact), $g_{nm}({\rm d})$
(Darwin), and $g_{nm}({\rm oo})$ (orbit--orbit).

The radiative rates ($A$-values) for electric dipole and quadrupole
transitions are respectively given in units of s$^{-1}$ by the expressions
\begin{equation}
A_{\rm E1}(k,i) = 2.6774\times 10^9(E_k-E_i)^3{1\over g_k}S_{\rm E1}(k,i)
\end{equation}
\begin{equation}
A_{\rm E2}(k,i) = 2.6733\times 10^3(E_k-E_i)^5{1\over g_k}S_{\rm E2}(k,i)
\end{equation}
where $S(k,i)$ is the line strength, $g_k$ is the statistical weight of the
upper level, and energies are in Rydberg units and lengths in Bohr radii.

Similarly for magnetic dipole and quadrupole transitions, the $A$-values are
respectively given by
\begin{equation}
A_{\rm M1}(k,i) = 3.5644\times 10^4(E_k-E_i)^3{1\over g_k}S_{\rm M1}(k,i)
\end{equation}
\begin{equation}
A_{\rm M2}(k,i) = 2.3727\times 10^{-2}(E_k-E_i)^5{1\over g_k}S_{\rm M2}(k,i)\ .
\end{equation}
Due to the strong magnetic interactions in this ion, the magnetic dipole
line strength is assumed to take the form
\begin{equation}
S_{\rm M1}(k,i)=|\langle|k|{\bf P}|i\rangle|^2
\end{equation}
where
\begin{equation} \label{m1op}
{\bf P}={\bf P^0}+{\bf P^1}=\sum_{n=1}^N
\{{\bf l}(n)+{\bf\sigma}(n)\}+{\bf P^{\rm rc}}\ .
\end{equation}
${\bf P^0}$ is the usual low-order M1 operator while
${\bf P^{\rm rc}}$ includes the relativistic corrections established
by \citet{dra71}.

Although the main astrophysical interest is for E1 K$\alpha$ decays, it is
shown here that some of the forbidden transitions display $A$-values comparable
with the E1 type and thus must be taken into account for accuracy. Furthermore,
in the case of the ${\rm 1s2s2p}\ ^4{\rm P}^{\rm o}_{5/2}$ state, radiative
decay can only occur through forbidden transitions.

\section{Numerical methods}

In the present work we employ three different computational packages to
study the properties of the $n=2$ vacancy states of the Li-like
Fe~{\sc xxiv}.

\subsection{\sc autostructure}
{\sc autostructure}, an extension by \citet{bad86,bad97} of the atomic
structure program
{\sc superstructure} \citep{eis74}, computes fine-structure
level energies, radiative and Auger rates in a Breit--Pauli
relativistic framework. Single electron orbitals, $P_{nl}(r)$, are
constructed by diagonalizing the non-relativistic Hamiltonian,
$H_{\rm nr}$, within a statistical Thomas--Fermi--Dirac model
potential $V(\lambda_{nl})$ \citep{eis69}. The $\lambda_{nl}$
scaling parameters are optimized variationally by minimizing a
weighted sum of the $LS$ term energies. $LS$ terms are
represented by configuration-interaction (CI) wavefunctions of
the type
\begin{equation}
\Psi=\sum_i c_i \phi_i\ .
\end{equation}
Continuum wavefunctions are constructed within the distorted-wave approximation.
Relativistic fine-structure levels and rates are obtained by diagonalizing
the Breit--Pauli Hamiltonian in intermediate coupling. The one- and two-body
operators---fine structure and non-fine structure (see Section 2)---have
been fully implemented to order $\alpha^2Z^4$ where $\alpha$ is the
fine-structure constant and $Z$ the atomic number. The relativistic corrections
to the M1 operator (see Eq.~\ref{m1op}) have been incorporated in
{\sc superstructure} by \citet{eis81}.

Fine tuning (semi-empirical corrections)---which is resourceful for
treating states that decay through
weak relativistic couplings (e.g. intercombination transitions)---takes the
form of term energy corrections (TEC). By considering the
relativistic wavefuntion, $\psi^{\rm r}_i$, in an perturbation expansion
of the non-relativistic functions $\psi^{\rm nr}_i$,
\begin{equation}
\psi^{\rm r}_i=\psi^{\rm nr}_i +\sum_{j\neq i}\psi^{\rm nr}_j\times
{\langle\psi^{\rm nr}_j|H_{\rm 1b}+H_{\rm 2b}|\psi^{\rm nr}_i\rangle\over
E^{\rm nr}_i-E^{\rm nr}_j}\ ,
\end{equation}
a modified $H_{nr}$ is constructed with improved estimates of the differences
$E^{\rm nr}_i-E^{\rm nr}_j$ so as to adjust the centers of gravity of the
spectroscopic terms
to the experimental values. This procedure therefore relies on the availability
of measured data.


\subsection{\sc hfr}
In {\sc hfr} \citep{cow81}, a set of orbitals are obtained
for each electronic configuration by solving the Hartree--Fock
equations for the spherically averaged atom. The equations are the
result of the application of the variational principle to the
configuration average energy. Relativistic corrections are also
included in this set of equations, i.e. the Blume--Watson
spin--orbit, mass-variation and one-body
Darwin terms. The Blume--Watson spin--orbit term comprises the part
of the Breit interaction that can be reduced to a one-body
operator.

The multiconfiguration Hamiltonian matrix is constructed and
diagonalized in the $LSJ\pi$ representation within the framework
of the Slater--Condon theory. Each matrix element is a sum of
products of Racah angular coefficients and radial integrals
(Slater and spin--orbit integrals), i.e.
\begin{equation}
\langle a|H|b\rangle = \sum_{i} c_{i}^{a,b} I_{i}^{a,b}\ .
\end{equation}
The radial parameters, $I_{i}^{a,b}$, can be adjusted to
fit the available experimental energy levels in a
least-squares approach. The eigenvalues and the
eigenstates obtained in this way ({\it ab initio} or
semi-empirically) are used to compute the wavelength and
oscillator strength for each possible transition.

Autoionization rates can be calculated using the perturbation
approach

\begin{equation}
  \begin{array}{ll}
    A^a & = \frac{2\pi}{\hbar} V_{\varepsilon}^2 \\
    & \\
     & = \frac{2\pi}{\hbar} |< \alpha LSJ\pi |
      H| \alpha' L'S'J' \varepsilon l \; LSJ\pi >|^2
  \end{array}
\end{equation}
where $\alpha$ summarizes the coupling scheme and the remaining
set of quantum numbers necessary to define the initial state, and
$\alpha'$ plays a similar role for the threshold state to which
the continuum electron, $\varepsilon l$, is coupled. The kinetic
energy of the free electron, $\varepsilon$, is determined as the
difference between the average energy of the autoionizing and the
threshold configurations. The radial wave functions of the initial
and final states are optimized separately. Both states are
calculated in intermediate coupling but CI
is accounted for only in the autoionizing states,
i.e. no interaction between threshold electronic configurations
is introduced.
The continuum orbitals, $P_{\varepsilon l}(r)$, are solutions of
the Hartree--Plus--Statistical-Exchange equations for fixed
positive values of the Lagrangian multipliers, $\varepsilon$
\citep{cow81}.


\subsection{\sc bprm}
The {\sc bprm} method is widely used in electron--ion scattering and in radiative
bound--bound and bound--free calculations. It is based of the close-coupling
approximation of
\citet{bursea71} whereby the wavefunctions for states
of an $N$-electron target and a colliding electron with total angular momentum
and parity $J\pi$ are expanded in terms of the target eigenfunctions
\begin{equation}\label{cc}
\Psi^{J\pi}={\cal A}\sum_i \chi_i{F_i(r)\over r}+\sum_jc_j\Phi_j\ .
\end{equation}
The functions $\chi_i$ are vector coupled products of the target eigenfunctions
and the angular components of the incident-electron functions, $F_i(r)$ are the
radial part of the latter and $\mathcal{A}$ is an antisymmetrization operator.
The functions $\Phi_j$ are bound-type functions of the total system constructed
with target orbitals; they are introduced to compensate for orthogonality
conditions imposed on the $F_i(r)$ and to improve short-range correlations.
The Kohn variational gives rise to a set of coupled integro-differential
equations that are solved by $R$-matrix techniques
\citep{bur71,ber74,ber78,ber87} within a box of radius, say,
$r\leq a$. In the asymptotic region ($r>a$) exchange between the outer
electron and the target ion can be neglected, and the wavefunctions can be then
approximated by Coulomb solutions. Resonance parameters are
obtained with the {\sc stgqb} module developed by \citet{qui96} and
\citet{qui98} whereby the resonance positions and widths are obtained from
fits of the eigenphase sum. Normalized partial
widths are defined from projections onto the open channels.

Breit--Pauli relativistic corrections have been introduced in the
$R$-matrix suite by \citet{sco80,sco82}, but the two-body terms
(see Eq.~\ref{2bt}) have not as yet been implemented.
Inter-channel coupling is equivalent to CI
in the atomic structure context, and thus the {\sc bprm} method
presents a formal and unified approach to
study the decay properties of both bound states and resonances.

\section{Radiation and Auger dampings}
When an electron or a photon are sufficiently energetic to excite a ground-state
ion to a K-vacancy resonance, the latter can either fluoresce or
autoionize (Auger decay). Illustrating these processes
with the resonances converging to the $n=2$ K thresholds in the
collisional excitation of Fe~{\sc xxiv}
and the photoexcitation of Fe~{\sc xxiii}, that is
\begin{eqnarray}
\left\{
\begin{array}{l}
{\rm Fe^{23+}}({\rm 1s}^2{\rm 2s}) + e^-\\
{\rm Fe^{22+}}({\rm 1s}^2{\rm 2s}^2) + h\nu
\end{array} \right\}
\rightarrow\left\{
\begin{array}{l}
{\rm Fe^{22+}}({\rm 1s2s}^2nl)\\
{\rm Fe^{22+}}({\rm 1s2s2p}nl)\\
{\rm Fe^{22+}}({\rm 1s2p}^2nl)
\end{array} \right\}\ ,
\end{eqnarray}
the decay manifold can be outlined as follows:
\begin{eqnarray}
\left\{
\begin{array}{l}
{\rm Fe^{22+}}({\rm 1s2s}^2nl)\\
{\rm Fe^{22+}}({\rm 1s2s2p}nl)\\
{\rm Fe^{22+}}({\rm 1s2p}^2nl)
\end{array} \right\}
 & \rightarrow & \left\{
 \begin{array}{l}
 {\rm Fe^{23+}}({\rm 1s2s}^2) + e^-\\
 {\rm Fe^{23+}}({\rm 1s2s2p}) + e^-\\
 {\rm Fe^{23+}}({\rm 1s2p}^2) + e^-
 \end{array} \right\} \label{direct} \\
 & \rightarrow & \left\{
 \begin{array}{l}
 {\rm Fe^{23+}}({\rm 1s}^2{\rm 2s}) + e^-\\
 {\rm Fe^{23+}}({\rm 1s}^2{\rm 2p}) + e^-\\
 \end{array} \right\} \label{KLn} \\
& \rightarrow & \left\{{\rm Fe^{23+}}({\rm 1s}^2nl) + e^-
\right\}\label{KLL}\\
& \rightarrow & \left\{
 \begin{array}{l}
     {\rm Fe^{22+}}({\rm 1s}^2{\rm 2s}^2) + h\nu\\
     {\rm Fe^{22+}}({\rm 1s}^2{\rm 2s2p}) + h\nu\\
     {\rm Fe^{22+}}({\rm 1s}^2{\rm 2p}^2) + h\nu
 \end{array} \right\} \label{Kn}\\
& \rightarrow & \left\{
 \begin{array}{l}
     {\rm Fe^{22+}}({\rm 1s}^2{\rm 2s}nl) + h\nu\\
     {\rm Fe^{22+}}({\rm 1s}^2{\rm 2p}nl) + h\nu
 \end{array} \right\}\ .\label{Ka}
\end{eqnarray}
The direct outer-shell ionization channels (Eq.~\ref{direct})
and the participator KL$n$ Auger channels (Eq.~\ref{KLn}) can be adequately
represented in the {\sc bprm} method by including in the
close-coupling expansion (\ref{cc}) configuration-states within the $n=2$
complex of the three-electron target. On the other hand,
in the KLL Auger process in Eq.~(\ref{KLL}), also
referred to as spectator Auger decay, the $nl$ Rydberg electron remains a
spectator. Its formal handling in the close-coupling approach is thus severely
limited to low-$n$ resonances as it implies the inclusion of target states
with $nl$ orbitals. Moreover,
it has been recently shown by \citet{pal02} that KLL
is the dominant Auger decay mode in the Fe sequence by no less than 75\%,
and leads to photoionization cross sections populated with
damped resonances of constant widths as $n\rightarrow\infty$
which causes the smearing of the edge.

Transitions in Eq.~(\ref{Kn}) and Eq.~(\ref{Ka}) lead to radiation damping. The
former, to be referred to as the K$n$ transition array, are driven by
the $n{\rm p}\rightarrow 1{\rm s}$ optical electron jump. The latter is the
K$\alpha$ transition array (${\rm 2p}\rightarrow 1{\rm s}$)
where again the $nl$ Rydberg electron remains a spectator; its dominant
width is therefore practically independent of $n$ \citep{pal02}.

The present treatment of Auger and radiative dampings within the {\sc bprm}
framework
uses the optical potential described by \cite{gorczyca96}
and \cite{gorczyca00}, where the resonance threshold energy acquires
an imaginary component. For example, the core energy of the
closed channel ${\rm 1s2s2p}nl$ is now expressed as
\begin{equation}
E_{1s^{-1}}\to E_{1s^{-1}} - i(\Gamma^a_{1s^{-1}} +\Gamma^r_{1s^{-1}})/2,
\end{equation}
where $\Gamma^a_{1s^{-1}}$ and $\Gamma^r_{1s^{-1}}$ are respectively
the Auger and
radiative widths of the 1s2s2p core. In the case of radiation damping,
the optical potential modifies the $R$-matrix to the complex form
\begin{equation}
R_{jj'}(E) = R^0_{jj'}(E) +2\sum_{nn'}d^0_{jn}d^0_{j'n'}(\gamma^{-1})_{nn'},
\end{equation}
where $R^0_{jj'}$ are the $R$-matrix elements without damping,
$d^0_{jn}$ are $(N+1)$-electron dipole matrix
elements and $\gamma^{-1}$ is a small inverted complex matrix
defined in Eq. (100) of \cite{robiche95}.

The calculations of collisional excitation and photoionization with
the {\sc bprm} method are carried out with
the standard $R$-matrix computer package of \cite{ber95} for the inner region
and on the asymptotic codes
{\sc stgfdamp} \citep{gorczyca96} and {\sc stgbf0famp} (Badnell,
unpublished) to determine cross sections including radiation
and Auger dampings.

\begin{singlespace}
\begin{table*}
\centering
\caption[]{Ion model key.
AST1--AST3: Present work ({\sc autostructure}).
HFR1--HFR3: Present work ({\sc hfr}).
BPR1: Present work ({\sc bprm}).
COR: Cornille data set from \citet{kat97}.
SAF: Safronova data set from \citet{kat97} and \citet{saf96}.
MCDF: Multiconfiguration Dirac--Fock calculation by \citet{chen86}.
}
\label{table1}
\begin{tabular}{llllllllllll}
\hline\hline
Feature& AST1& AST2& AST3& HFR1& HFR2& HFR3& BPR1& COR& SAF& MCDF\\
\hline
Orthogonal orbital basis   & Yes& Yes& Yes& Yes & No & No & Yes& Yes& Yes &Yes \\
CI from $n>2$ complexes    & No & No & Yes& No  & No & Yes& No & No & Yes &Yes \\
Breit interaction          & No & Yes& Yes& Yes & Yes& Yes& No & No & Yes &Yes \\
QED effects                & No & No & No & No  & No & No & No & No & Yes &Yes \\
Semi-empirical corrections & No & No & Yes& No  & No & Yes& No & No & No  & No \\
\hline
\end{tabular}
\end{table*}
\end{singlespace}

\section{Ion models}

Since the present study of the Fe Li-like system has been approached as a test
case, the atomic data
are computed with several ion models and extensively compared with other
data sets. This methodology is destined to bring out the dominant physical effects
and the flaws and virtues of the different numerical packages.
Additionally, it provides statistics for determining accuracy ratings,
something which has not been fully established in the past.
The main features of each approximation are summarized in the key in
Table~\ref{table1}.

Three calculations with {\sc autostructure} are listed: AST1,
the ion is modeled with states from configurations within the $n=2$
complex and excludes the Breit interaction, i.e. the
relativistic two-body operators in Eq.~(\ref{2bt}); AST2, the
same as AST1 but takes into account the Breit interaction; AST3 includes
the latter, single and double excitations to the $n=3$ complex
and TEC. AST3 allows the evaluation of CI effects
from higher complexes and to fine-tune the data for accuracy. Orthogonal
orbital bases are generated for each of these three approximations
by minimizing the
sum of the energies of all the $LS$ terms comprising the respective
ion representations.
A dilemma quickly arises in {\sc autostructure} calculations
regarding the ion model in the context of Auger processes,
whether to use Li-type orbitals (parent ion) or those
of the He-like remnant. By comparing with results from the more formal
{\sc bprm} method, it becomes clear that the latter type is the superior choice.
On the other hand, the situation is less certain for the
K$\alpha$ radiative data due to the absence of noticeable differences. In
this case, and due to better agreement with previous work, the $A$-values
have been calculated with parent orbitals.

Three computations with {\sc hfr} are discussed: HFR1 is
equivalent to AST2 as the ion model with states within the $n=2$
complex with an orthogonal orbital basis. The 1s and 2s orbitals
are obtained by minimizing the energy of the ${\rm 1s}^2{\rm 2s}$
term whereas the 2p is optimized with ${\rm 1s}^2{\rm 2p}$. HFR2
employs the ion model of HFR1 but with non-orthogonal orbital
bases generated for each configuration by minimizing their average
energy. Comparisons of HFR1 and HFR2 will thus give estimates of
core relaxation effects (CRE) which have been long known
\citep{how78a,how78b,bre79} but generally neglected in the more
recent work on the Fe isonuclear sequence. In HFR3 non-orthogonal
bases are used, full $n=3$ CI is taken into account and the
radial integrals are fitted to reproduce experimental energies
(this approximation should then be comparable to AST3). BPR1 is
a  computation with {\sc bprm} wherein the He-like target is
represented with the 19 levels from the 1s$^2$, 1s2s, and 1s2p
configurations. Since {\sc bprm} does not take into account the
Breit interaction, BPR1 should be comparable with AST1.

We also compare with three external data sets (see Table~\ref{table1}).
COR, corresponds to the data set
referred to as ``Cornille" in \citet{kat97} computed with the program
{\sc autolsj} \citep{dub81}, an earlier but similar implementation of
{\sc autostructure}. SAF contains the data set ``Safronova" in \citet{kat97}
and energy levels reported in \citet{saf96} that have been obtained with a
$1/Z$ perturbation method. This method uses a hydrogenic orbital basis, the
correlation energy includes contributions from both discrete and continuum
states, and the two-body operators of the Breit interaction and QED effects
are obtained in a hydrogenic approximation through screening constants.
MCDF \citep{chen86} contains data computed in a
multiconfiguration Dirac--Fock method that accounts for the Breit interaction
and QED in the transition energy, but
excludes the exchange interaction between the bound and continuum electrons.

In our comparisons two external computations are excluded.
\citet{lem84} have computed Auger rates
with {\sc hfr} in a single configuration approximation (i.e. no CI
even within $n=2$), the Breit interaction is not taken into account
and the Coulomb integrals are empirically scaled by 15\% to allow for
neglected effects. The large discrepancies found with our {\sc hfr} calculations
can be perhaps attributed their questionable atomic model.
\citet{nah01} have computed with {\sc bprm} radiative and Auger widths
for the 1s2s2p states. There is good general accord with our BPR1 results, and
since they only report a reduced data set, it will not be further
discussed.

\section{Energies and wavelengths}
In Table~\ref{table2} we compare present level energies with
experiment and SAF. It may be seen that the energies obtained for
the K-vacancy levels with approximation AST1 are on average
$10\pm 2$~eV higher than experiment. By including the Breit
interaction (AST2), and mainly due to the contribution from the
non-fine structure two-body terms, this discrepancy is slightly
reduced to $8\pm 1$~eV. Further consideration of CI, i.e. from
configurations of the $n=3$ complex, does not bring about
noticeable improvements. Results obtained with BPR1 bear a
similar degree of discord with measured values. This
systematic difference is partly
due to neglected interactions (e.g. QED), but also to the fact
that orthogonal orbital bases are used to represent the
ground and lowly excited bound states, in the one hand, and the
highly excited K-vacancy resonances on the other thus discarding
CRE. This assertion is supported by a comparison of average differences
of HFR1 (excludes CRE) and HFR2 (includes CRE) with experiment:
$5\pm 1$~eV  and $2\pm 1$~eV respectively.
Fine tuning, invoked in approximations AST3 and HFR3, results in
theoretical levels within 1~eV of experiment, comparable to the
accuracy of 1.5~eV displayed by SAF. For the unobserved ${\rm
1s2s2p}\ ^4{\rm P}^{\rm o}_{5/2}$ level, an energy of 6.6285(3)
keV is predicted which is in good accord with value of 6.6283 keV
quoted by SAF.

In Table~\ref{table3}  we compare line wavelengths derived from
the AST3 and HFR3 approximations with experiment and other
theoretical results. The measurements were made by \citet{bei93}
with a high-resolution Bragg crystal spectrometer on the
Princeton Large Torus Tokamak. Our previous criticism regarding
the incompleteness of the experimental data sets can be appreciated
in this comparison. With respect to experiment,
differences with HFR3 and SAF are not larger than 0.4~m\AA\ while
those with AST3 and MCDF are within 0.6~m\AA\ and 0.8~m\AA\,
respectively. This level of accord is somewhat outside of the
average experimental precision of 0.23~m\AA. The values listed by
COR are systematically shorter than experiment by $\sim 3$~m\AA.
In general, differences between the AST3, HFR3,
SAF and MCDF data sets show scatters with standard deviations
not larger than 0.3~m\AA\ which can perhaps be taken as a lower
bound of the theoretical accuracy.


\section{Radiative rates}
A Li-like K-vacancy state decays radiatively by emitting a K$\alpha$ photon
\begin{equation}
{\rm 1s2s}^{n_k}{\rm 2p}^{m_k}\ ^{(2S_k+1)}L_{J_k}\rightarrow
{\rm 1s}^22l_i\ ^2L'_{J_i} + \lambda_{{\rm K}\alpha}
\end{equation}
where the strong transitions are the dipole spin-allowed ($2S_k+1=2$), but
intercombination transitions ($2S_k+1=4$) can also take place via
subtle relativistic couplings. Furthermore, we hereby demonstrate that in some
cases the forbidden transitions cannot be put aside.

\begin{figure}
\resizebox{\hsize}{!}{\includegraphics{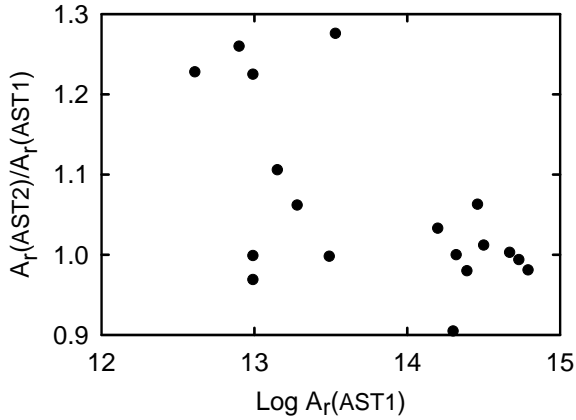}}
\caption{Comparison of $A$-values (s$^{-1}$) for K transitions
in Fe~{\sc xxiv} computed with approximations AST1 and
AST2. Differences are due to Breit interaction.}
\label{fig1}
\end{figure}

\begin{figure}
\resizebox{\hsize}{!}{\includegraphics{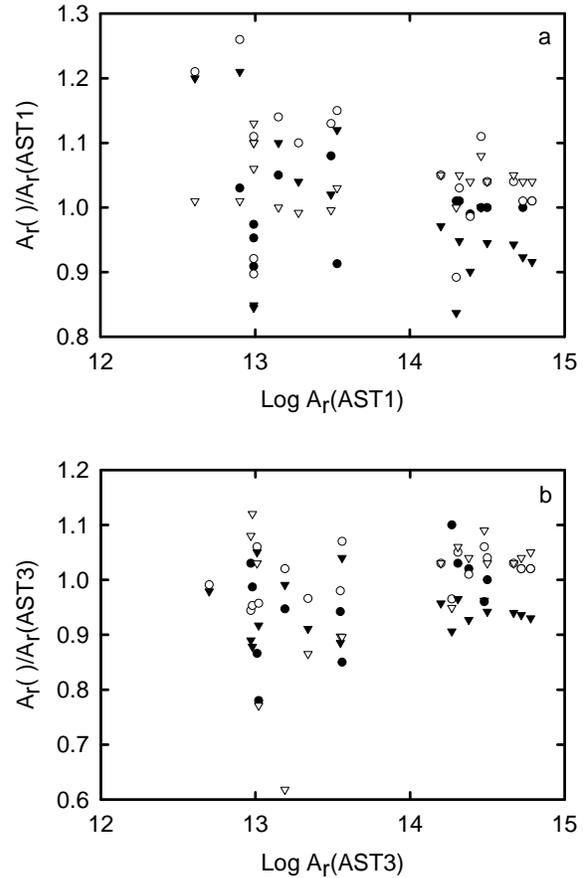}}
\caption{Comparison of {\sc autostructure} $A$-values (s$^{-1}$) for
K transitions in Fe~{\sc xxiv} with other approximations and external
data sets. (a) AST1 with: HFR2 (triangles);
COR (filled circles); SAF (circles); and MCDF (filled triangles). (b)
AST3 with: HFR3 (triangles); COR (filled circles); SAF (circles); and MCDF
(filled triangles).}
\label{fig2}
\end{figure}

In Table~\ref{table4} we present transition probabilities computed in the
different approximations together with those from previous work
(COR, SAF, and MCDF). In the following discussion, we exclude the transitions
10--3, 12--1, 13--2, and 18--2 as they are severely affected by cancellation
and nothing further can be asserted about their radiative properties.
In Fig.~\ref{fig1}
we compare $A$-values computed in AST2 with those in AST1 where significant
differences are found. In general, the inclusion of the Breit interaction
(AST2) increases rates; while the variations are not larger than 10\%
for the spin allowed transitions that exhibit large rates
($\log A_{\rm r}>14$), the enhancement in the intersystem transitions
(5--1, 6--1, and 13--3) can be as large as 25\%. Inclusion of CI
from the $n=3$ complex leads to changes not larger than 2\%, but
the fitting with TEC, as expected, causes differences mostly in the
sensitive intersystem transitions. By comparing HFR1 and HFR2
(see Table~\ref{table4}), it can
be concluded that CRE tend to increase $A$-values but seldom by more than 10\%;
the exceptions are the transitions affected by strong cancellation
effects (e.g. 12--1, 13--2).

In Fig.~\ref{fig2}a the transition probabilities computed in
approximation AST1 are compared with those by HFR2, COR, SAF and MCDF.
While there is as expected excellent agreement with COR (within 10\%),
the data in HFR2 and SAF are on average higher by $\sim 5\%$ with scatters of
$\pm 4$\% and $\pm 12$\%, respectively. Differences with MCDF are as large
as 21\%. The discord with HFR2 is due to CRE while that
with SAF and MCDF is believed to be due to the contributions of the
relativistic two-body corrections excluded in AST1.
This assertion is supported by a further comparison
with the data in AST3 (Fig.~\ref{fig2}b); now the agreement with
SAF and MCDF has improved to $\sim 10$\% while discrepancies as
large as 25\% are found with COR where the Breit interaction was neglected.
The larger differences now
found with HFR3 (15\%) are an indication that the Blume--Watson
screening in {\sc hfr} does not account adequately for the Breit
interaction. The outcome of this comparison clearly brings out the
relevance of relativistic effects in the radiative decay, and give us
confidence on the accuracy ranking ($\sim 10$\%) that can be assigned to the
$A$-values in AST3 which we regard our best.

We have found that the K-vacancy states in Li-like iron, in addition to their
dipole allowed manifold, can also decay radiatively
via unusually strong magnetic
transitions. As shown in Table~\ref{table5}, the $A$-values for the M2
components in 10--3 and 13--2 are almost as large as their E1
counterparts, and therefore must be taken into account in order to maintain
accuracy. The situation becomes critical for the
${\rm 1s2s2p}\ ^4{\rm P}^{\rm o}_{5/2}$ metastable
which is shown to decay through both
M1 and M2 transitions (see Table~\ref{table5}). It
may be also appreciated that the M1 $A$-value must be calculated with the
relativistically corrected operator (see Eq.~\ref{m1op}) since the
difference with the uncorrected version is 5 orders of magnitude.
\cite{chen81} have assumed that this state decays radiatively only via the M2
transition, and quote a value of $A_{\rm r}=6.57\times 10^9$~s$^{-1}$ in
good agreement (7\%) with the present $A_{\rm M2}=6.16\times 10^9$~s$^{-1}$.

\begin{figure}
\resizebox{\hsize}{!}{\includegraphics{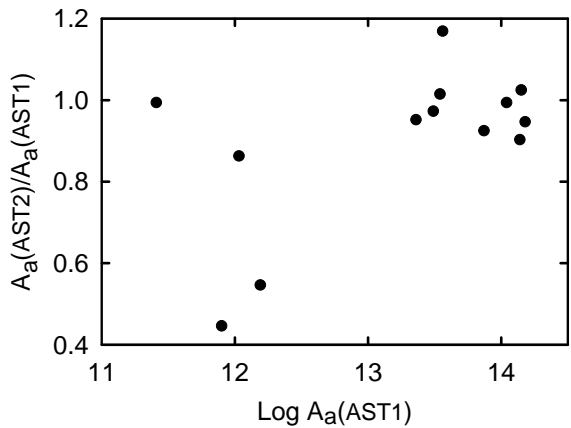}}
\caption{Comparison of Auger rates (s$^{-1}$) for K-vacancy
levels of Fe~{\sc xxiv} computed with approximations AST1 and
AST2. Differences are due to Breit interaction.}
\label{fig3}
\end{figure}

\section{Auger rates}

While the radiative transition probabilities can be resolved satisfactorily,
the effects of the magnetic couplings and CRE on the Auger rates are
more evident and thus larger the discrepancies. A Li-like $1s2l2l'$ level
autoionizes through the reaction
\begin{equation}
{\rm 1s2s}^{n_k}{\rm 2p}^{m_k}\ ^{(2S_k+1)}L_{J_k}\rightarrow
{\rm 1s}^2\ ^2{\rm S}_0 + {\rm e}^-
\end{equation}
that ends up in the ground state of the He-like child ion. A
comparison of rates is given in Table~\ref{table6}. As before,
due to strong cancellation effects, we exclude the ${\rm
1s}(^2{\rm S}){\rm 2s2p}(^3{\rm P}^{\rm o})\ ^2{\rm P}^{\rm
o}_{3/2}$ and $^4{\rm P}^{\rm o}_{1/2}$ states from further
discussion. By comparing data from approximations AST1 and AST2
(see Fig.~\ref{fig3}), it is found significant sensitivity to the
Breit interaction: states with $\log A_{\rm a}>13$ are in general
reduced by no more than 11\%, but the smaller values show
decrements as large as a factor of 2. As shown in
Table~\ref{table7}, the spin--spin interaction can cause drastic
changes in the rates, not only due to level coupling within the
parent bound configurations (bound--bound coupling) but also
involving the final continuum configuration (bound--free
coupling). An outstanding illustration of this correlation is the
${\rm 1s2s2p}\ ^4{\rm P}^{\rm o}_{5/2}$ state which can only
autoionize through the spin--spin interaction. By contrast, CI
from the $n=3$ complex is found to be relatively unimportant, but
the TEC lead to noticeable changes (25\%) in the quartet states,
e.g. ${\rm 1s2p}^2\ ^4{\rm P}_J$, that can only decay through
relativistic intersystem couplings that are sensitive to level
separation. The good agreement ($\sim 10$\%) between AST1 and
BPR1 for states with $\log A_{\rm a}>13$ (see Table~\ref{table6})
reinforces the {\sc autostructure} numerical formulation of
autoionization processes. CRE in Auger decay are disclosed in the
comparison of HFR1 and HFR2 where it is found that relaxation
generally increase widths by ($11\pm 5$)\%. Discrepancies between
AST2 and HFR2 and AST3 and HFR3, which can be as large as 45\%
for transitions with $\log A_{\rm a}>13$, are believed to be due
to both CRE and the oversimplified implementation of the Breit
interaction in {\sc hfr}.

\begin{figure}
\resizebox{\hsize}{!}{\includegraphics{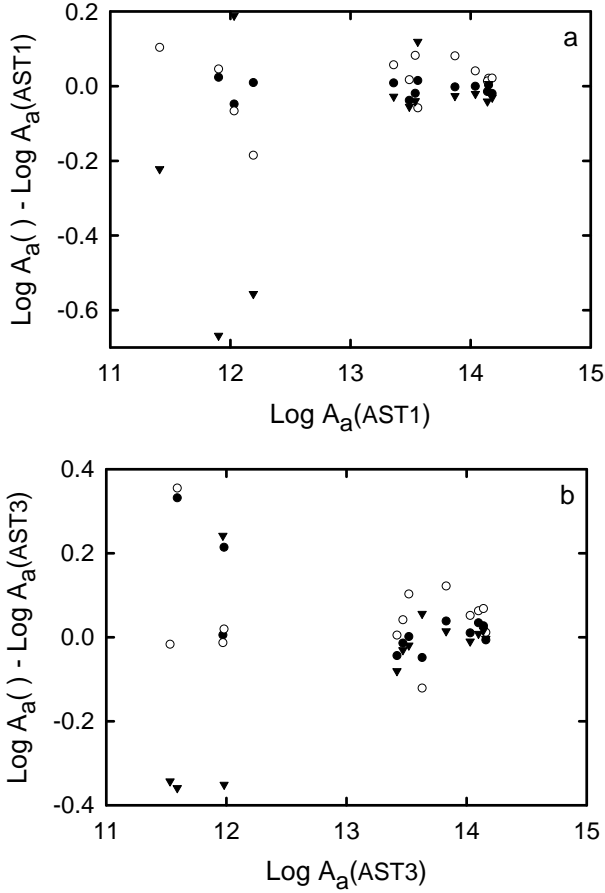}}
\caption{Comparison of {\sc autostructure} Auger rates (s$^{-1}$) for
K-vacancy levels in Fe~{\sc xxiv} with previous data sets. (a) AST1 with: COR
(filled circles); SAF (circles); and MCDF (filled triangles). (b) AST3 with:
COR (filled circles); SAF (circles); and MCDF (filled triangles).}
\label{fig4}
\end{figure}

In Fig.~\ref{fig4} Auger rates in AST1 and AST3 are compared with COR,
SAF, and MCDF.
While agreement between COR and AST1 is within 10\%, it clearly deteriorates
with AST3; this is further evidence of the neglect of the Breit interaction
by COR. Significant differences are also found
with SAF and MCDF in particular for the smaller values ($\log A_{\rm a}< 13$).
Focusing our discussion on the larger rates, data by SAF are on average
8\% higher than AST1 which is a worrying outcome as the inclusion of the Breit
interaction in general decreases our rates thus magnifying the discrepancy.
This can be appreciated in the comparison of SAF with AST3 in Fig.~\ref{fig4}b where
the larger differences are found for decays subject to strong spin--spin
bound--free correlation (see Table~\ref{table7}), and can perhaps be attributed to
its deficient treatment in the SAF approach. By contrast, the
discord between AST1 and MCDF for the larger rates (up to 32\%) is
reduced to within 15\% when the Breit interaction is taken into account.

The lack of data stability for Auger transitions with $\log A_{\rm a}< 13$
is further put in evidence in the tricky decay of the
${\rm 1s2s2p}\ ^4{\rm P}^{\rm o}_{5/2}$ state. While there is good
agreement with \citet{chen81} for the dominant radiative M2 $A$-value
(see Section~7), their Auger rate of $6.53\times 10^9$~s$^{-1}$ is a
factor of 3 larger thus predicting a lower fluorescence yield (0.50)
than the present (0.76) for this state.

\section{$B_{\rm r}$ and $Q_{\rm d}$ factors}

In the spectral synthesis of dielectronic satellite lines, relevant
parameters for a $k\rightarrow i$ radiative emission are the branching ratio
\begin{equation}
B_{\rm r}(k,i)\equiv {A_{\rm r}(k,i)\over A_{\rm r}(k)+A_{\rm a}(k)}
\end{equation}
and the satellite intensity factor
\begin{equation}
Q_{\rm d}(k,i)\equiv g_kB_{\rm r}(k,i)A_{\rm a}(k)
\end{equation}
where $A_{\rm r}(k,i)$, $A_{\rm r}(k)=\sum_i A_{\rm r}(k,i)$,
$A_{\rm a}(k)$, and $g_k$ are respectively the $A$-value, total
radiative width, Auger rate and statistical weight of the upper
$k$ level. In Table~\ref{table8} we compare our best data set
(AST3) with COR, SAF, and MCDF. For $B_{\rm r}>0.1$, the
agreement is within 5\% except for the COR 13--3 and the SAF
11--1 lines where it deteriorates to 9\%. The former, being an
intercombination transition, is sensitive to the atomic model
while level 11 is subject to admixture. For $B_{\rm r}<0.1$, the
accord is within 15\% if transitions affected with cancellation
are put aside. For $Q_{\rm d}>10^{13}$~s$^{-1}$, agreement with
COR, SAF, and MCDF is respectively within 10\%, 25\%, and 15\%,
but for the smaller values, discrepancies up to a factor of 9 do
appear.

\begin{figure*}
\centering
\includegraphics[width=12cm]{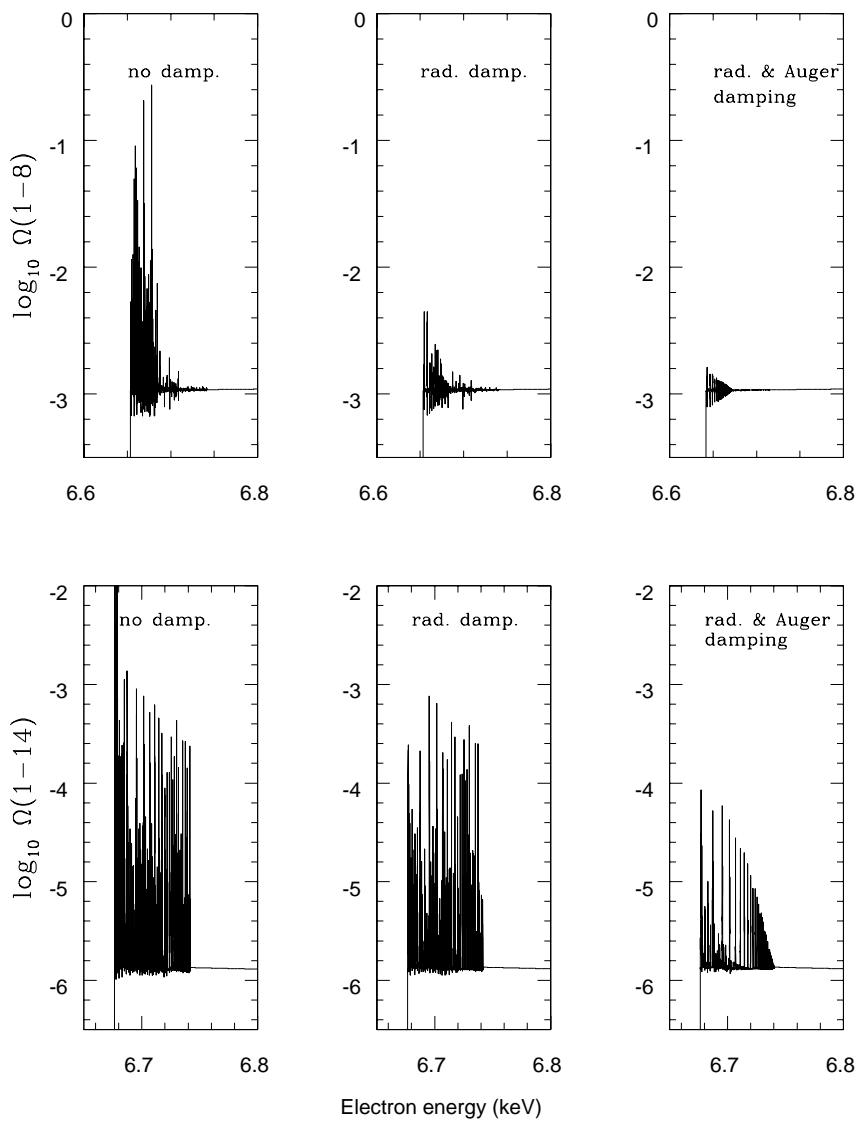}
\caption{Comparison of electron impact
collision strengths for K-shell excitation in Fe~{\sc xxiv}
computed with the {\sc bprm} method. The left panels depict
collision strengths for the 1--8 and 1--14 transitions computed
without damping. The effects of radiation and spectator Auger dampings
can be appreciated in the middle and right panels, respectively.}
\label{fig5}
\end{figure*}

\section{Electron impact inner-shell excitation of Fe~{\sc xxiv}}

Collision strengths for the electron impact excitation of the
${\rm 1s}^2{\rm 2s}$ and ${\rm 1s}^2{\rm 2p}$ states to
${\rm 1s}2l2l'$ of Fe~{\sc xxiv} have been computed with the {\sc bprm}
method. The target representation includes only the 19 levels within
the $n=2$ complex since exploratory calculations with $n=3$ target states
lead to negligible differences. We are particularly concerned with the
effects of radiative and Auger dampings and the convergence of the
partial wave expansion.

In Fig.~\ref{fig5} collision strengths for both an allowed (1--8) and
a forbidden (1--14) transition are shown. Although the background cross
section is generally small ($\log \Omega <-2$), specially for the latter type,
they both display dense resonance structures in the region just
above threshold that rise by several orders of magnitude.  When radiation
damping is introduced, however, resonances are washed out in
the allowed transition and significantly attenuated in the forbidden case,
trend that is further completed when Auger damping is taken into account.
In agreement with \citet{whi02}, the effect of the combined dampings on
the low-temperature effective collision strengths can be drastic as
illustrated in Table~\ref{table9} where differences of factors are seen.
The extreme case is the forbidden transition
1--13 that is overestimated by nearly two orders of magnitude if damping is
altogether neglected and by a factor of two with the exclusion of Auger
damping. It must be pointed out that the calculation by \citet{ball01}
of inner-shell excitation of Li- and Be-like Fe
does not take into account Auger damping.

With regards to relativistic effects, the collision strengths for the fine
structure transitions have been calculated in three different
approximations: (a) $LS$-coupling followed by
algebraic recoupling; (b) $LS$-coupling
followed by recoupling with term coupling coefficients that account for
target fine structure and (c) the relativistic Hamiltonian (Eq.~\ref{1bt})
that includes only the one-body operators.
Good agreement is found between approximations (b) and (c) while large
discrepancies are found with (a). These results indicate that relativistic
effects must be taken into account in the scattering formulation and that
the two-body corrections, which are not implemented in {\sc bprm}, are small and
can be neglected in this case.

Under coronal ionization conditions the temperatures of maximum
abundance of Fe~{\sc xxiii} and Fe~{\sc xxiv} are $\sim 2\times
10^7$ K  and $\sim 4\times 10^7$ K respectively; effective
collision strengths must be then computed at temperatures of up
to $10^8$K. To ensure accuracy in the Maxwellian averaging
integral, collision strengths are computed in a range up to 4000
Ryd where partial wave convergence becomes the main issue.  The
calculation is performed in two stages: a full {\sc bprm}
calculation for total angular momentum of the ($N+1$)-electron
system in the range $0\le J \le 10$ and a non-exchange
calculation for higher $J$ which is carried out in $LS$ coupling
and then recoupled with term coupling coefficients. Very good
agreement is found with the Coulomb--Born--Exchange collision
strengths by \cite{goett84} for transitions from the ground state
in the non-resonant region.

Maxwellian averaged collision strengths are listed in Table~\ref{table10}
for the electron-temperature range $5\leq \log T\leq 8$ for all the $n=2$
K transitions. Infinite-temperature limits are also tabulated which
for allowed transitions are $\Omega(\infty)=4gf/\Delta E$---where
$gf$ and $\Delta E$ are respectively
the weighted oscillator strength and excitation energy for the
transition---and $\Omega(\infty)=\Omega_{\rm CB}$ for forbidden transitions
with $\Omega_{\rm CB}$ being the Coulomb--Born high-energy limit.
The $gf$ and $\Omega_{\rm CB}$ have been computed with {\sc autostructure}
with approximation AST1. Good agreement (within 10\%) is found in the
entire range with
both the Coulomb--Born--Exchange results of \citet{goett84} for transitions
from the ground level and data set computed with the $R$-matrix by
\citet{whi02} using a more elaborate target ($n\leq5$ complexes). This
is the result of the general irrelevance of resonances caused
by the damping processes.

\begin{figure}
\resizebox{\hsize}{!}{\includegraphics{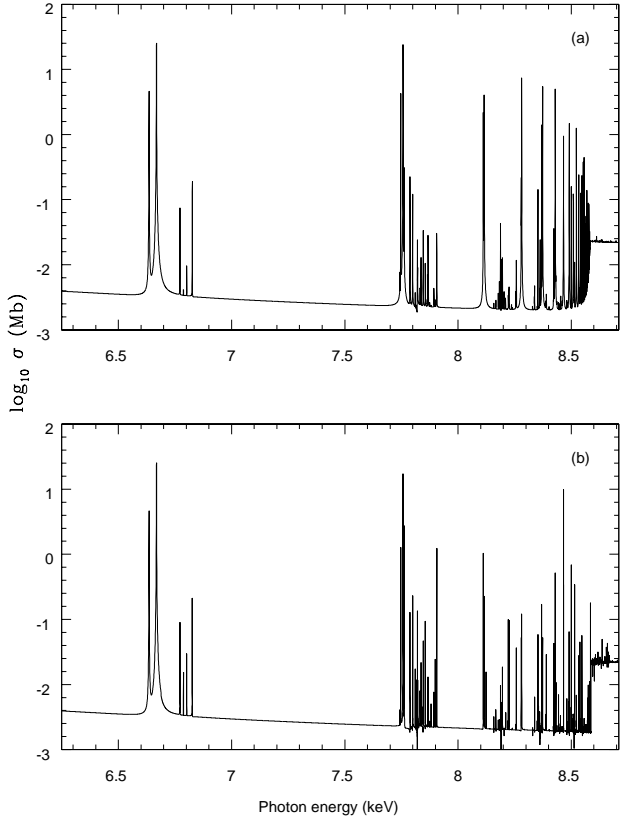}}
\caption{Photoabsorption cross section of Fe~{\sc xxiii}. The upper
panel (a) depicts the cross section computed including radiative and
spectator-Auger damping effects. The lower panel (b) shows the
same cross section when these effects are neglected.}
\label{fig6}
\end{figure}

\begin{figure}
\resizebox{\hsize}{!}{\includegraphics{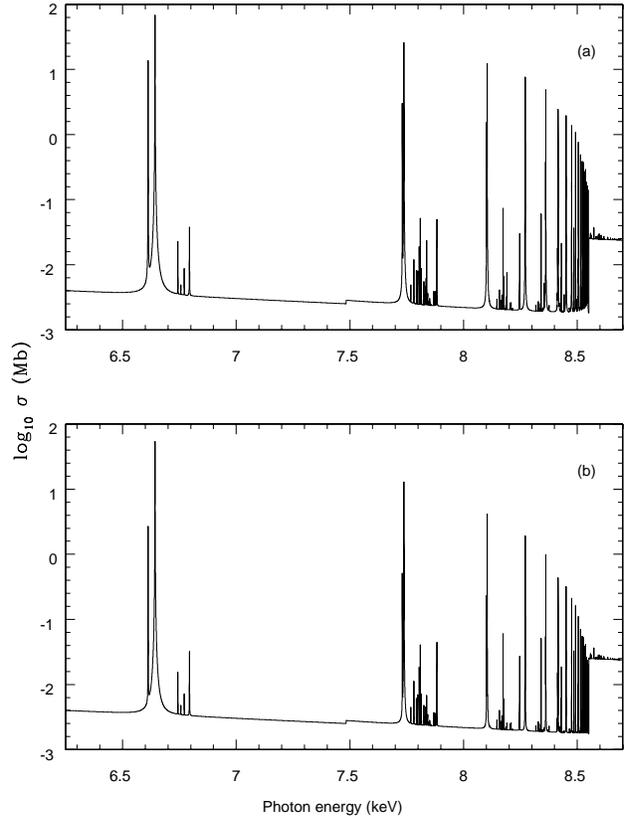}}
\caption{Comparison between the (a) photoabsorption cross section
and the (b) photoionization cross section computed with
{\sc autostructure} assuming Lorentzian resonance profiles.}
\label{fig7}
\end{figure}

\section{Inner-shell photoabsorption and photoionization of Fe~{\sc xxiii}}

The inner-shell photoabsorption cross section of the Fe~{\sc
xxiii} ground states has been computed with {\sc bprm} using the
same 19-level Li-like target model described in Section~10. As
shown in Fig.~\ref{fig6}a, the cross section is dominated by a
series of symmetric resonances of constant width that cause the
smearing of the K edge. This unusual resonance behavior, as
explained by \citet{pal02}, is a consequence of the dominance of
K$\alpha$ and KLL dampings. When such damping is neglected (see
Fig.~\ref{fig6}b), only the lowest $n=2$ resonance array is
accurately represented with the present $n=2$ target model
whereas the widths of the higher components are markedly
underestimated and decrease with $n$ maintaining edge sharpness.

A further key point to make is that when damping is fully taken into account
the inner-shell photoabsorption and photoionization processes must be treated
separately. In the former, the integrated cross section under the resonance
must remain constant in spite of the broadening caused by damping
so as to conserve oscillator strength. In the latter, the cross section
is actually reduced since radiation damping leads to radiative de-excitation
instead of photoionization. Unfortunately,
there is as yet no formal procedure to separate the radiative de-excitation
component in {\sc bprm}.

An alternative method is to compute photoabsorption and photoionization cross
sections with {\sc autostructure} by estimating a central-field background
cross section and making use of the isolated resonance approximation to
compute resonance positions, radiative decay rates and Auger widths
for all levels with configurations
${\rm 1s}2l2l'nl''$. Assuming Lorentzian profiles, resonances in
photoabsorption and photoionization cross sections
can be approximated by the expressions
\begin{equation}
\sigma^{\rm abs}= { gf (\Gamma_{\rm r}+\Gamma_{\rm a})
\over (E-E_{\rm c})^2 + 1/4(\Gamma_{\rm r}+\Gamma_{\rm a})^2}
\end{equation}
and
\begin{equation}
\sigma^{ion}= { gf\Gamma_{\rm a}
\over (E-E_{\rm c})^2 + 1/4(\Gamma_{\rm r}+\Gamma_{\rm a})^2}\ ,
\end{equation}
where $gf$ is the weighted absorption oscillator strength,$\Gamma_{\rm r}$ and $\Gamma_{\rm a}$
are respectively the radiative and Auger widths, and $E$ and $E_c$
the photon and resonance energies. In Fig.~\ref{fig7} the photoabsorption
and photoionization cross sections calculated with {\sc autostructure} are
depicted. The attenuated resonance heights in the photoionization can be
appreciated (see Fig.~\ref{fig7}b),
and a good quantitative resemblance is found for the former with
that obtained with {\sc bprm} (Fig.~\ref{fig6}a).

\begin{figure*}
\centering
\includegraphics[width=12cm]{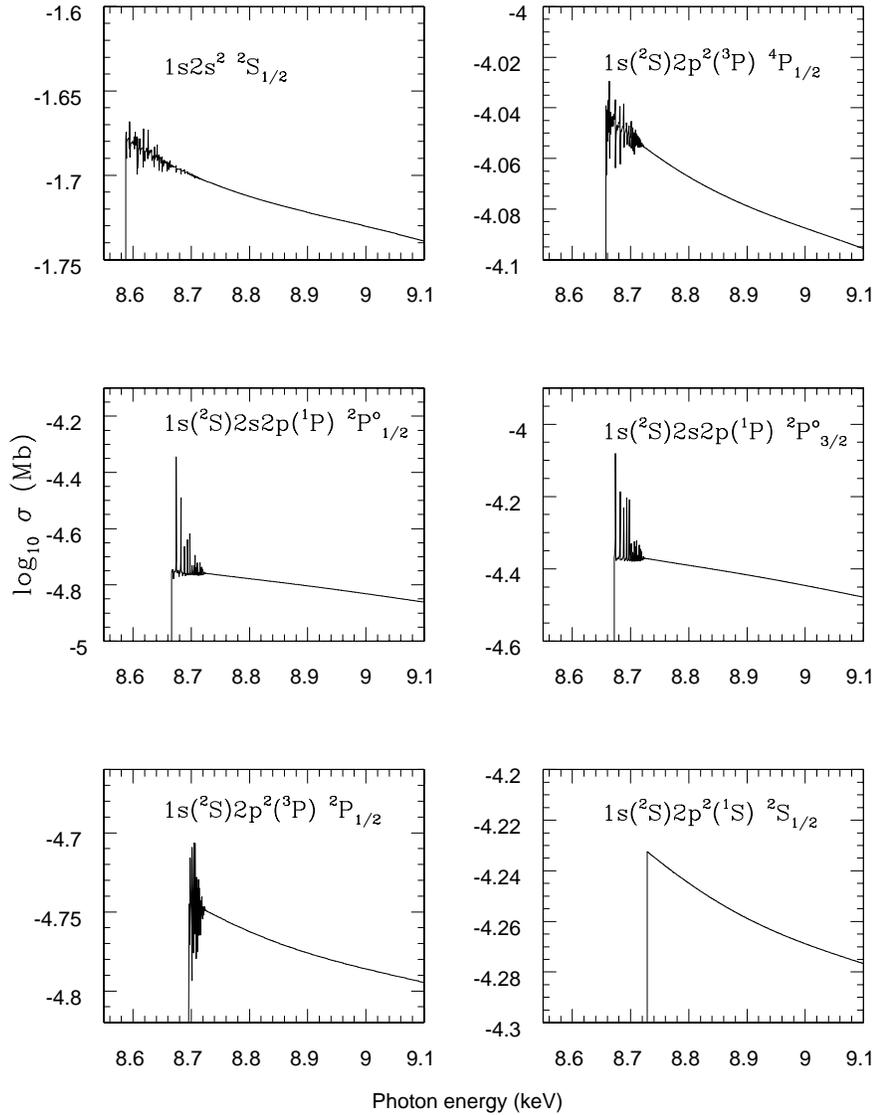}
\caption{Partial photoionization cross sections from the ground level
of Fe~{\sc xxiii} leaving Fe~{\sc xxiv} in a K-vacancy state.}
\label{fig8}
\end{figure*}

Partial photoionization cross sections of the Fe~{\sc xxiii}
ground state leaving the Li-like remnant in a K vacancy state are
displayed in Fig.~\ref{fig8}. Only the stronger transitions are
included where it is seen that the transition to the ${\rm
1s2s}^2~^2{\rm S}_{1/2}$ level dominates. Since the radiative
transition rates for this state are an order of magnitude lower
than its Auger width (see Tables~\ref{table4} and \ref{table6}),
the most probable final state in its decay tree is the ground
state of Fe~{\sc xxv}. Therefore, the inner-shell photoionization
of the ground state of Fe~{\sc xxiii} yields a double ionization
rather than a satellite line. Furthermore, since the ${\rm
1s}^2{\rm 2s2p}\ ^3{\rm P}^{\rm o}_0$ and $^3{\rm P}^{\rm o}_2$
excited states of Fe~{\sc xxiii} are metastable, their
photoionization contribution should be in principle included in
models. However, unlike the ground state, their photoionization
leaves the ion in K levels with strong radiative channels that
produce satellite lines.

\section{Summary and conclusions}

As a start in a project to compute improved atomic data for the spectral
modeling of Fe K lines, we have carried out extensive calculations and
comparisons of atomic data for modeling of the K spectrum of Li-like
Fe~{\sc xxiv}. The data set includes energy levels, radiative and Auger rates,
collision strengths, and total and partial photoionization cross sections.
Primary aims have been to select an applicable computational
platform and an efficient strategy to generate accurate and complete data
sets for other ions of the first row of the Fe isonuclear sequence.

We have studied several physical effects, namely orbital
representations, core relaxation,
CI, relativistic corrections, cancellation,
semi-empirical corrections, and the damping of resonances by radiative and
spectator Auger decay. For an $N$-electron ion, we have found that the
most realistic representation is to have different orbital
bases for the K-vacancy states, on the one hand, and for the
valence states of the $N$- and $(N-1)$-electron systems on the
other. This is available in {\sc hfr}, but most other
codes use orthogonal orbital bases for computational efficiency.
In the case the {\sc autostructure}, which uses a
distorted-wave approach to compute Auger rates, orbitals of the
$(N-1)$-electron system must then be used. Core relaxation leads to
increases in the radiative and Auger widths no larger than 10\%.

Level coupling within
the $n=2$ complex has been found to be key, thus seriously
questioning the reliability of the atomic model adopted by
\citet{lem84}. CI from higher complexes contributes negligibly.
Contributions from the two-body relativistic operators, both fine
structure and non-fine structure, play a conspicuous role in the
decay of K-vacancy states of this ion, particularly in the Auger
pathways. Electron correlation could be then interpreted as
being highly magnetic: bound--free spin--spin effects have been
shown to be important within the $n=2$ complex and specially critical
for the Auger decay of the metastable ${\rm 1s2s2p}\ ^4{\rm
P}^{\rm o}_{5/2}$ state. This state is also shown to decay
radiatively through forbidden M1 and M2 transitions, the former
requiring a relativistic corrected transition operator to avoid
errors in the line strength of several orders of magnitude. In
this highly ionized magnetic scenario, computer programs that do
not include a formal numerical implementation of the Breit
interaction, or neglect it, have limited applicability. Such
is the case of {\sc bprm} and {\sc hfr}. Some of the large discrepancies
found for the smallest rates have been attributed to strong cancellation
effects. Fine tuning has
been found to be a useful option to attain high numerical
accuracy, particularly for line identification and to render
intersystem couplings that can be very sensitive to level
separations.

In the light of the problems discussed above, none of the codes seems
to be the platform of choice for the calculation of radiative and Auger rates.
We therefore employ several computational platforms to
treat inner-shell processes which has proven to be key in elucidating the
physics involved, and has been used previously by COR and SAF
and more recently by \citet{sav02}. This approach has therefore been retained
in our current calculations of other members of the Fe isonuclear sequence.

The present {\sc autostructure} calculations are an independent
validation and refinement of that performed in COR;
the level of agreement found at the different stages confirms this
assertion. The excellent accord also obtained with the radiative rates by
SAF allows us to establish a firm ranking of 10\% for the present $A$-values.
On the other hand, the fairly large discrepancies with the SAF Auger rates
are believed to be caused by their approximate treatment of the Breit
interaction in terms of screening constants. We therefore
rank the present autoionization data at better than 15\%. We can also conclude
by comparing with SAF that the attained precision for the
K-vacancy level energies of  $\pm 4$ eV is a representative lower bound for
current numerical methods. This however implies fine tuning that relies on
spectroscopic measurements. Since complete experimental level structures
are not available for most systems, further experiments would be welcome.

Radiative and spectator Auger dampings have been found to be of
fundamental importance in the calculation of K-shell
photoionization and electron excitation processes. In the former,
resonances converging to the K threshold acquire a peculiar
behavior that leads to edge smearing which, as discussed by
\citet{pal02}, has diagnostic potential in astrophysical
plasmas. With regards to the latter, resonances are practically
washed out, thus simplifying target modeling or the choice of a
suitable numerical approach. This assertion is supported by the
good agreement (10\%) of the present excitation rates with the
Coulomb--Born--Exchange results of \citet{goett84} and with those
in $R$-matrix calculation by \citet{whi02} who used a more
refined target. We have also found that the ground state of
Fe~{\sc xxiii} is mainly photoionized to the ${\rm 1s2s}^2~^2{\rm
S}_{1/2}$ K level of Fe~{\sc xxiv} which rapidly autoionizes
rather than fluoresces. Thus  K$\alpha$ emission from a Fe
Li-like ion is mainly the result of electron impact excitation
and dielectronic recombination.


\begin{acknowledgements}
We are indebted to Dr. Nigel Badnell from the University of Strathclyde,
UK, for invaluable discussions regarding the {\sc autostructure} options,
Auger processes in general and
the peculiar decay properties of the K-vacancy metastable state of this ion.
Also to Dr. Marguerite Cornille, Observatoire de Meudon, France, for
details about the COR and SAF calculations.
CM acknowledges a Senior Research Associateship from the National Research
Council, and MAB support from FONACIT, Venezuela, under contract No.
S1-20011000912. Support for this research was provided in part by a grant
through the NASA Astrophysics Theory Program.
\end{acknowledgements}


\begin{singlespace}
\begin{table*}[p]
\centering
\caption[]{Comparison of level energies (keV) for the $n=2$ complex
of Fe~{\sc xxiv} (see approximation key in Table~\ref{table1}).
Experimental values from \citet{shi00}.}
\label{table2}
\begin{tabular}{rllllllllll}
\hline\hline
$i$& State & Expt & AST1 & AST2 & AST3 & HFR1 & HFR2 & HFR3 & BPR1 & SAF\\ \hline
1  & ${\rm 1s}^2{\rm 2s}\ ^2{\rm S}_{1/2}$                                      &0.0      &0.0      &0.0     &0.0     &0.0     &0.0     &0.0     &0.0    &0.0     \\
2  & ${\rm 1s}^2{\rm 2p}\ ^2{\rm P}^{\rm o}_{1/2}$                              &0.04860  &0.04801  &0.04928 &0.04778 &0.04843 &0.04850 &0.04860 &       &0.04854\\
3  & ${\rm 1s}^2{\rm 2p}\ ^2{\rm P}^{\rm o}_{3/2}$                              &0.06457  &0.06696  &0.06689 &0.06498 &0.06446 &0.06454 &0.06457 &       &0.06453\\
4  & ${\rm 1s}{\rm 2s}^2\ ^2{\rm S}_{1/2}$                                      &6.6004   &6.6099   &6.6070  &6.6003  &6.6051  &6.6018  &6.6004  &6.6072 &6.6011  \\
5  & ${\rm 1s}(^2{\rm S}){\rm 2s2p}(^3{\rm P}^{\rm o})\ ^4{\rm P}^{\rm o}_{1/2}$&6.6137   &6.6202   &6.6189  &6.6131  &6.6175  &6.6129  &6.6131  &6.6177 &6.6135  \\
6  & ${\rm 1s}(^2{\rm S}){\rm 2s2p}(^3{\rm P}^{\rm o})\ ^4{\rm P}^{\rm o}_{3/2}$&6.6167   &6.6253   &6.6227  &6.6169  &6.6221  &6.6178  &6.6173  &6.6230 &6.6171  \\
7  & ${\rm 1s}(^2{\rm S}){\rm 2s2p}(^3{\rm P}^{\rm o})\ ^4{\rm P}^{\rm o}_{5/2}$&         &6.6376   &6.6342  &6.6285  &6.6330  &6.6295  &6.6265  &       &6.6283  \\
8  & ${\rm 1s}(^2{\rm S}){\rm 2s2p}(^3{\rm P}^{\rm o})\ ^2{\rm P}^{\rm o}_{1/2}$&6.6535   &6.6624   &6.6598  &6.6525  &6.6567  &6.6538  &6.6537  &6.6605 &6.6534  \\
9  & ${\rm 1s}(^2{\rm S}){\rm 2s2p}(^3{\rm P}^{\rm o})\ ^2{\rm P}^{\rm o}_{3/2}$&6.6619   &6.6732   &6.6697  &6.6623  &6.6665  &6.6641  &6.6618  &6.6708 &6.6624  \\
10 & ${\rm 1s}(^2{\rm S}){\rm 2p}^2(^3{\rm P})\ ^4{\rm P}_{1/2}$                &6.6710   &6.6781   &6.6770  &6.6706  &6.6753  &6.6709  &6.6708  &6.6764 &6.6717  \\
11 & ${\rm 1s}(^2{\rm S}){\rm 2s2p}(^1{\rm P}^{\rm o})\ ^2{\rm P}^{\rm o}_{1/2}$&6.6764   &6.6866   &6.6841  &6.6764  &6.6814  &6.6784  &6.6766  &6.6831 &6.6765  \\
12 & ${\rm 1s}(^2{\rm S}){\rm 2s2p}(^1{\rm P}^{\rm o})\ ^2{\rm P}^{\rm o}_{3/2}$&6.6792   &6.6896   &6.6867  &6.6791  &6.6839  &6.6812  &6.6790  &6.6869 &6.6795  \\
13 & ${\rm 1s}(^2{\rm S}){\rm 2p}^2(^3{\rm P})\ ^4{\rm P}_{3/2}$                &6.6793   &6.6868   &6.6855  &6.6792  &6.6829  &6.6790  &6.6786  &6.6853 &6.6798  \\
14 & ${\rm 1s}(^2{\rm S}){\rm 2p}^2(^3{\rm P})\ ^4{\rm P}_{5/2}$                &6.6850   &6.6946   &6.6917  &6.6850  &6.6900  &6.6865  &6.6857  &6.6932 &6.6856  \\
15 & ${\rm 1s}(^2{\rm S}){\rm 2p}^2(^1{\rm D})\ ^2{\rm D}_{3/2}$                &6.7027   &6.7137   &6.7118  &6.7027  &6.7082  &6.7050  &6.7029  &6.7112 &6.7042  \\
16 & ${\rm 1s}(^2{\rm S}){\rm 2p}^2(^3{\rm P})\ ^2{\rm P}_{1/2}$                &6.7046   &6.7159   &6.7128  &6.7041  &6.7099  &6.7068  &6.7048  &6.7141 &6.7052  \\
17 & ${\rm 1s}(^2{\rm S}){\rm 2p}^2(^1{\rm D})\ ^2{\rm D}_{5/2}$                &6.7090   &6.7211   &6.7176  &6.7089  &6.7147  &6.7120  &6.7096  &6.7189 &6.7097  \\
18 & ${\rm 1s}(^2{\rm S}){\rm 2p}^2(^3{\rm P})\ ^2{\rm P}_{3/2}$                &6.7224   &6.7349   &6.7315  &6.7225  &6.7268  &6.7247  &6.7219  &6.7329 &6.7230  \\
19 & ${\rm 1s}(^2{\rm S}){\rm 2p}^2(^1{\rm S})\ ^2{\rm S}_{1/2}$                &6.7415   &6.7541   &6.7514  &6.7414  &6.7468  &6.7448  &6.7412  &6.7519 &6.7428  \\
\hline
\end{tabular}
\end{table*}

\begin{table*}[p]
\centering
\caption[]{Comparison of wavelengths (\AA) for K transitions in Fe~{\sc xxiv}
(see approximation key in Table~\ref{table1}).
Transition labels from \citet{see86} and
tokamak measurements (uncertainties in brackets) by \citet{bei93}.}
\label{table3}
\begin{tabular}{crrllllll}
\hline\hline
Label & $k$ & $i$ & Expt & AST3 & HFR3 & COR & SAF & MCDF \\ \hline
p &  4 & 2 &1.89219(25) & 1.8922 & 1.8924 & 1.8894 & 1.8924 & 1.8927 \\
o &  4 & 3 &1.89680(20) & 1.8971 & 1.8970 & 1.8946 & 1.8969 & 1.8973 \\
v &  5 & 1 &            & 1.8748 & 1.8748 &        & 1.8748 & 1.8752 \\
u &  6 & 1 &1.87347(35) & 1.8737 & 1.8736 & 1.8712 & 1.8738 & 1.8742 \\
  &  7 & 1 &            & 1.8706 &        &        &        &        \\
  &  7 & 3 &            & 1.8890 &        &        &        &        \\
r &  8 & 1 &1.86325(20) & 1.8639 & 1.8634 & 1.8611 & 1.8635 & 1.8640 \\
q &  9 & 1 &1.86104(15) & 1.8610 & 1.8611 &        & 1.8610 & 1.8613 \\
i & 10 & 2 &            & 1.8720 & 1.8722 &        & 1.8722 & 1.8725 \\
h & 10 & 3 &            & 1.8768 & 1.8768 &        & 1.8766 & 1.8771 \\
t & 11 & 1 &1.85693(20) & 1.8568 & 1.8570 & 1.8543 & 1.8571 & 1.8571 \\
s & 12 & 1 &            & 1.8563 & 1.8563 & 1.8535 & 1.8563 & 1.8564 \\
g & 13 & 2 &            & 1.8697 & 1.8701 &        & 1.8699 & 1.8702 \\
f & 13 & 3 &            & 1.8745 & 1.8746 & 1.8724 & 1.8743 & 1.8747 \\
e & 14 & 3 &1.87246(35) & 1.8729 & 1.8726 & 1.8703 & 1.8727 & 1.8730 \\
k & 15 & 2 &1.86325(20) & 1.8630 & 1.8632 & 1.8601 & 1.8630 & 1.8631 \\
l & 15 & 3 &            & 1.8677 & 1.8677 & 1.8652 & 1.8674 & 1.8676 \\
d & 16 & 2 &            & 1.8626 & 1.8627 & 1.8594 & 1.8628 & 1.8629 \\
c & 16 & 3 &            & 1.8674 & 1.8672 &        & 1.8672 & 1.8673 \\
j & 17 & 3 &1.86576(12) & 1.8661 & 1.8658 & 1.8631 & 1.8659 & 1.8660 \\
b & 18 & 2 &            & 1.8576 & 1.8579 & 1.8542 & 1.8578 & 1.8578 \\
a & 18 & 3 &1.86207(30) & 1.8623 & 1.8624 & 1.8593 & 1.8622 & 1.8622 \\
n & 19 & 2 &            & 1.8523 & 1.8526 & 1.8488 & 1.8523 & 1.8521 \\
m & 19 & 3 &1.85693(20) & 1.8570 & 1.8570 & 1.8539 & 1.8566 & 1.8565 \\
\hline
\end{tabular}
\end{table*}

\begin{table*}[p]
\centering
\caption[]{Comparison of $A$-values ($10^{13}$~s$^{-1}$) for K transitions
in Fe~{\sc xxiv} (see approximation key in Table~\ref{table1}).
Transition labels from \citet{see86}. Note: $a\pm b\equiv a\times 10^{\pm b}$.}
\label{table4}
\begin{tabular}{crrlllllllll}
\hline\hline
Label& $k$& $i$& AST1& AST2& AST3& HFR1& HFR2& HFR3& COR& SAF & MCDF\\ \hline
p & 4 & 2  & 9.76$-$1 & 9.46$-$1 & 9.27$-$1 & 9.62$-$1 & 1.03$+$0 & 1.00$+$0 & 9.51$-$1& 8.75$-$1& 8.25$-$1\\
o & 4 & 3  & 9.85$-$1 & 9.84$-$1 & 9.52$-$1 & 1.04$+$0 & 1.08$+$0 & 1.07$+$0 & 9.39$-$1& 9.07$-$1& 8.36$-$1\\
v & 5 & 1  & 4.06$-$1 & 4.98$-$1 & 4.97$-$1 & 3.72$-$1 & 4.08$-$1 & 2.92$-$1 &         & 4.92$-$1& 4.86$-$1\\
u & 6 & 1  & 1.40$+$0 & 1.55$+$0 & 1.55$+$0 & 1.26$+$0 & 1.40$+$0 & 9.60$-$1 & 1.47$+$0& 1.59$+$0& 1.54$+$0\\
  & 7 & 1  & 6.18$-$4 & 6.18$-$4 & 6.16$-$4 &          &          &          &         &         &         \\
  & 7 & 3  & 1.93$-$5 & 1.94$-$5 & 1.94$-$5 &          &          &          &         &         &         \\
r & 8 & 1  & 2.88$+$1 & 3.06$+$1 & 3.01$+$1 & 3.04$+$1 & 3.10$+$1 & 3.29$+$1 & 2.88$+$1& 3.19$+$1& 2.89$+$1\\
q & 9 & 1  & 4.70$+$1 & 4.71$+$1 & 4.71$+$1 & 4.72$+$1 & 4.94$+$1 & 4.86$+$1 &         & 4.87$+$1& 4.43$+$1\\
i & 10 & 2 & 1.90$+$0 & 2.02$+$0 & 2.17$+$0 & 1.72$+$0 & 1.89$+$0 & 1.88$+$0 &         & 2.10$+$0& 1.98$+$0\\
h & 10 & 3 & 1.77$-$2 & 7.70$-$3 & 9.12$-$3 & 1.68$-$2 & 1.79$-$2 & 1.60$-$2 &         & 9.30$-$3& 1.27$-$2\\
t & 11 & 1 & 2.01$+$1 & 1.82$+$1 & 1.86$+$1 & 1.87$+$1 & 2.01$+$1 & 1.76$+$1 & 2.03$+$1& 1.79$+$1& 1.68$+$1\\
s & 12 & 1 & 8.92$-$1 & 5.90$-$1 & 4.19$-$1 & 1.05$+$0 & 6.57$-$1 & 1.25$+$0 & 4.41$-$1& 7.78$-$2& 3.23$-$1\\
g & 13 & 2 & 6.21$-$2 & 6.63$-$3 & 4.51$-$3 & 7.77$-$3 & 9.03$-$3 & 1.06$-$2 &         & 2.40$-$3& 3.42$-$3\\
f & 13 & 3 & 8.01$-$1 & 1.01$+$0 & 1.06$+$0 & 7.54$-$1 & 8.11$-$1 & 8.13$-$1 & 8.23$-$1& 1.01$+$0& 9.67$-$1\\
e & 14 & 3 & 3.11$+$0 & 3.11$+$0 & 3.58$+$0 & 2.81$+$0 & 3.10$+$0 & 3.21$+$0 & 3.37$+$0& 3.51$+$0& 3.17$+$0\\
k & 15 & 2 & 3.13$+$1 & 3.17$+$1 & 3.14$+$1 & 3.14$+$1 & 3.26$+$1 & 3.24$+$1 & 3.15$+$1& 3.27$+$1& 2.96$+$1\\
l & 15 & 3 & 3.39$+$0 & 4.32$+$0 & 3.64$+$0 & 3.37$+$0 & 3.49$+$0 & 3.26$+$0 & 3.09$+$0& 3.90$+$0& 3.80$+$0\\
d & 16 & 2 & 5.39$+$1 & 5.35$+$1 & 5.31$+$1 & 5.40$+$1 & 5.62$+$1 & 5.53$+$1 & 5.39$+$1& 5.44$+$1& 4.97$+$1\\
c & 16 & 3 & 1.58$+$1 & 1.63$+$1 & 1.60$+$1 & 1.62$+$1 & 1.66$+$1 & 1.65$+$1 &         & 1.65$+$1& 1.53$+$1\\
j & 17 & 3 & 2.09$+$1 & 2.09$+$1 & 2.05$+$1 & 2.12$+$1 & 2.19$+$1 & 2.17$+$1 & 2.11$+$1& 2.16$+$1& 1.98$+$1\\
b & 18 & 2 & 1.15$+$0 & 7.70$-$1 & 9.69$-$1 & 1.16$+$0 & 1.21$+$0 & 1.24$+$0 & 1.25$+$0& 8.63$-$1& 7.57$-$1\\
a & 18 & 3 & 6.16$+$1 & 6.04$+$1 & 6.07$+$1 & 6.18$+$1 & 6.43$+$1 & 6.37$+$1 & 6.20$+$1& 6.21$+$1& 5.64$+$1\\
n & 19 & 2 & 9.78$-$1 & 1.20$+$0 & 1.03$+$0 & 1.18$+$0 & 1.11$+$0 & 1.06$+$0 & 8.89$-$1& 1.09$+$0& 1.08$+$0\\
m & 19 & 3 & 2.46$+$1 & 2.42$+$1 & 2.40$+$1 & 2.43$+$1 & 2.56$+$1 & 2.49$+$1 & 2.44$+$1& 2.43$+$1& 2.22$+$1\\
\hline
\end{tabular}
\end{table*}

 \begin{table*}[p]
 \centering
 \caption[]{$A$-values ($10^{9}$~s$^{-1}$) for K transitions with sizable
 magnetic components computed in approximation AST3. E1: electric dipole.
 M2: magnetic quadrupole. M1: magnetic dipole. M1*: magnetic dipole computed
 with uncorrected operator. Note: $a\pm b\equiv a\times 10^{\pm b}$.}
 \label{table5}
 \begin{tabular}{rrllll}
 \hline\hline
 $k$ & $i$ & E1 & M2& M1 & M1* \\ \hline
  7& 1 & 0.0      & 6.16$+$0 & 0.0      &          \\
  7& 3 & 0.0      & 0.0      & 1.94$-$1 & 6.11$-$7 \\
 10& 3 & 9.07$+$1 & 5.04$-$1 & 0.0      &          \\
 13& 2 & 3.99$+$1 & 5.19$+$0 & 0.0      &          \\
 \hline
 \end{tabular}
\end{table*}

\begin{table*}[p]
\centering
\caption[]{Comparison of Auger rates ($10^{13}$~s$^{-1}$) for K-vacancy
states in Fe~{\sc xxiv} (see approximation key in Table~\ref{table1}). Note:
$a\pm b\equiv a\times 10^{\pm b}$}
\label{table6}
\begin{tabular}{rllllllllll}
\hline\hline
$i$ &AST1 &AST2 &AST3 &HFR1 &HFR2 &HFR3 &BPR1& COR& SAF& MCDF\\ \hline
 4& 1.40$+$1& 1.44$+$1& 1.43$+$1& 1.25$+$1& 1.34$+$1& 1.34$+$1& 1.45$+$1& 1.41$+$1& 1.47$+$1&1.42$+$1\\
 5& 1.88$-$2& 1.45$-$3& 1.33$-$3& 1.36$-$2& 1.54$-$2& 1.09$-$2& 1.57$-$2&         & 1.19$-$2&5.57$-$3\\
 6& 7.96$-$2& 3.55$-$2& 3.91$-$2& 5.74$-$2& 6.56$-$2& 4.31$-$2& 7.07$-$2& 8.40$-$2& 8.85$-$2&1.71$-$2\\
 7& 0.00$+$0& 1.99$-$4& 1.97$-$4& 0.00$+$0& 0.00$+$0& 0.00$+$0& 0.00$+$0&         &         &        \\
 8& 3.67$+$0& 4.29$+$0& 4.24$+$0& 2.94$+$0& 3.42$+$0& 2.92$+$0& 3.87$+$0& 3.80$+$0& 3.21$+$0&4.83$+$0\\
 9& 8.99$-$4& 2.34$-$2& 1.41$-$2& 5.01$-$2& 3.01$-$2& 8.57$-$2& 1.55$-$2&         & 3.02$-$2&5.74$-$2\\
10& 2.55$-$2& 2.53$-$2& 3.37$-$2& 1.58$-$2& 1.94$-$2& 2.25$-$2& 3.15$-$2&         & 3.24$-$2&1.53$-$2\\
11& 7.43$+$0& 6.87$+$0& 6.77$+$0& 6.91$+$0& 7.16$+$0& 7.55$+$0& 7.74$+$0& 7.40$+$0& 8.96$+$0&7.00$+$0\\
12& 1.10$+$1& 1.10$+$1& 1.07$+$1& 9.77$+$0& 1.05$+$1& 1.04$+$1& 1.11$+$1& 1.10$+$1& 1.21$+$1&1.05$+$1\\
13& 1.55$-$1& 8.44$-$2& 9.66$-$2& 1.18$-$1& 1.37$-$1& 1.41$-$1& 1.78$-$1& 1.58$-$1& 1.01$-$1&4.30$-$2\\
14& 2.31$+$0& 2.20$+$0& 2.61$+$0& 1.75$+$0& 2.05$+$0& 2.12$+$0& 2.56$+$0& 2.36$+$0& 2.64$+$0&2.17$+$0\\
15& 1.39$+$1& 1.26$+$1& 1.25$+$1& 1.17$+$1& 1.29$+$1& 1.30$+$1& 1.38$+$1& 1.35$+$1& 1.44$+$1&1.27$+$1\\
16& 1.06$-$1& 9.16$-$2& 9.39$-$2& 6.60$-$2& 8.17$-$2& 8.11$-$2& 7.01$-$2& 9.50$-$2& 9.08$-$2&1.64$-$1\\
17& 1.52$+$1& 1.44$+$1& 1.37$+$1& 1.31$+$1& 1.44$+$1& 1.43$+$1& 1.47$+$1& 1.46$+$1& 1.60$+$1&1.42$+$1\\
18& 3.44$+$0& 3.49$+$0& 3.28$+$0& 3.05$+$0& 3.37$+$0& 3.27$+$0& 3.19$+$0& 3.29$+$0& 4.16$+$0&3.14$+$0\\
19& 3.09$+$0& 3.00$+$0& 2.92$+$0& 2.40$+$0& 2.77$+$0& 2.76$+$0& 2.75$+$0& 2.83$+$0& 3.21$+$0&2.72$+$0\\
\hline
\end{tabular}
\end{table*}

\begin{table*}[p]
\centering
\caption[]{Spin--spin contribution to Auger rates ($10^{13}$~s$^{-1}$).
SS: bound--free spin--spin coupling neglected. SS*: bound--free spin--spin
coupling included. Note: $a\pm b\equiv a\times 10^{\pm b}$}
\label{table7}
\begin{tabular}{rlll}
\hline\hline
$i$ & AST1 & AST1+SS & AST1+SS* \\ \hline
  4 & 1.40$+$1 & 1.31$+$1 & 1.40$+$1\\
  5 & 1.88$-$2 & 3.70$-$3 & 3.42$-$3\\
  6 & 7.96$-$2 & 4.27$-$2 & 2.96$-$2\\
  7 & 0.0      & 0.0      & 1.99$-$4\\
  8 & 3.67$+$0 & 3.92$+$0 & 3.98$+$0\\
  9 & 8.99$-$4 & 1.61$-$1 & 4.24$-$3\\
 10 & 2.55$-$2 & 2.11$-$2 & 2.69$-$2\\
 11 & 7.43$+$0 & 6.47$+$0 & 7.52$+$0\\
 12 & 1.10$+$1 & 1.02$+$1 & 1.09$+$1\\
 13 & 1.55$-$1 & 7.99$-$2 & 3.82$-$2\\
 14 & 2.31$+$0 & 2.00$+$0 & 2.06$+$0\\
 15 & 1.39$+$1 & 1.14$+$1 & 1.41$+$1\\
 16 & 1.06$-$1 & 7.37$-$2 & 1.01$-$1\\
 17 & 1.52$+$1 & 1.29$+$1 & 1.57$+$1\\
 18 & 3.44$+$0 & 3.42$+$0 & 3.11$+$0\\
 19 & 3.09$+$0 & 2.67$+$0 & 3.09$+$0\\
\hline
\end{tabular}
\end{table*}

\begin{table*}[p]
\centering
\caption[]{Comparison of radiative branching ratios $B_{\rm r}$ and
satellite intensity $Q_{\rm d}$ factors (see approximation key in
Table~\ref{table1}). Transition labels from \citet{see86}.
Note: $a\pm b\equiv a\times 10^{\pm b}$.}
\label{table8}
\begin{tabular}{crrlllllllllll}
\hline\hline
 & & &\multicolumn{2}{c}{AST3}& & \multicolumn{2}{c}{COR}& &
 \multicolumn{2}{c}{SAF} & &\multicolumn{2}{c}{MCDF}\\
\cline{4-5}\cline{7-8}\cline{10-11}\cline{13-14}\\
Label & $k$ & $i$ & $B_{\rm r}(k,i)$ & $Q_{\rm d}(k,i)$ & &
$B_{\rm r}(k,i)$ & $Q_{\rm d}(k,i)$ & &
$B_{\rm r}(k,i)$ & $Q_{\rm d}(k,i)$ & &
$B_{\rm r}(k,i)$ & $Q_{\rm d}(k,i)$ \\
 & & & & ($10^{13}$ s$^{-1}$) & & & ($10^{13}$ s$^{-1}$) & &
 & ($10^{13}$ s$^{-1}$) & & & ($10^{13}$ s$^{-1}$) \\ \hline
p &  4& 2& 5.72$-$2&  1.64$+$0& &  6.00$-$2&  1.68$+$0& &  5.29$-$2&  1.56$+$0& &5.20$-$2 & 1.48$+$0\\
o &  4& 3& 5.88$-$2&  1.68$+$0& &  5.90$-$2&  1.66$+$0& &  5.49$-$2&  1.62$+$0& &5.25$-$2 & 1.50$+$0\\
v &  5& 1& 9.97$-$1&  2.66$-$3& &          &          & &  9.76$-$1&  2.32$-$2& &9.90$-$1 & 1.10$-$2\\
u &  6& 1& 9.75$-$1&  1.53$-$1& &  9.46$-$1&  3.17$-$1& &  9.47$-$1&  3.35$-$1& &9.90$-$1 & 6.78$-$2\\
  &  7& 1& 7.40$-$1&  8.76$-$4& &          &          & &          &          & &         &         \\
  &  7& 3& 2.32$-$2&  2.75$-$5& &          &          & &          &          & &         &         \\
r &  8& 1& 8.76$-$1&  7.44$+$0& &  8.83$-$1&  6.72$+$0& &  9.09$-$1&  5.83$+$0& &8.55$-$1 & 8.28$+$0\\
q &  9& 1& 1.00$+$0&  5.64$-$2& &          &          & &  9.99$-$1&  1.21$-$1& &9.98$-$1 & 2.29$-$1\\
i & 10& 2& 9.81$-$1&  6.61$-$2& &          &          & &  9.81$-$1&  6.35$-$2& &9.85$-$1 & 3.01$-$2\\
h & 10& 3& 4.12$-$3&  2.77$-$4& &          &          & &  4.35$-$3&  2.82$-$4& &6.30$-$3 & 1.93$-$4\\
t & 11& 1& 7.33$-$1&  9.92$+$0& &  7.33$-$1&  1.08$+$1& &  6.67$-$1&  1.19$+$1& &7.05$-$1 & 9.88$+$0\\
s & 12& 1& 3.76$-$2&  1.61$+$0& &  3.80$-$2&  1.70$+$0& &  6.41$-$3&  3.09$-$1& &3.00$-$2 & 1.25$+$0\\
g & 13& 2& 3.90$-$3&  1.51$-$3& &          &          & &  2.16$-$3&  8.72$-$4& &3.38$-$3 & 5.81$-$4\\
f & 13& 3& 9.13$-$1&  3.53$-$1& &  8.27$-$1&  5.23$-$1& &  9.07$-$1&  3.67$-$1& &9.53$-$1 & 1.64$-$1\\
e & 14& 3& 5.78$-$1&  9.06$+$0& &  5.88$-$1&  8.34$+$0& &  5.71$-$1&  9.04$+$0& &5.93$-$1 & 7.72$+$0\\
k & 15& 2& 6.61$-$1&  3.29$+$1& &  6.55$-$1&  3.53$+$1& &  6.41$-$1&  3.70$+$1& &6.43$-$1 & 3.25$+$1\\
l & 15& 3& 7.66$-$2&  3.82$+$0& &  6.40$-$2&  3.47$+$0& &  7.64$-$2&  4.41$+$0& &8.25$-$2 & 4.18$+$0\\
d & 16& 2& 7.68$-$1&  1.44$-$1& &  7.72$-$1&  1.47$-$1& &  7.67$-$1&  1.39$-$1& &7.65$-$1 & 2.51$-$1\\
c & 16& 3& 2.31$-$1&  4.34$-$2& &          &          & &  2.32$-$1&  4.21$-$2& &2.35$-$1 & 7.70$-$2\\
j & 17& 3& 6.00$-$1&  4.92$+$1& &  5.92$-$1&  5.17$+$1& &  5.73$-$1&  5.52$+$1& &5.83$-$1 & 4.95$+$1\\
b & 18& 2& 1.49$-$2&  1.96$-$1& &  1.90$-$2&  2.47$-$1& &  1.29$-$2&  2.14$-$1& &1.26$-$2 & 1.58$-$1\\
a & 18& 3& 9.35$-$1&  1.23$+$1& &  9.31$-$1&  1.23$+$1& &  9.25$-$1&  1.54$+$1& &9.35$-$1 & 1.18$+$1\\
n & 19& 2& 3.68$-$2&  2.15$-$1& &  3.20$-$2&  1.79$-$1& &  3.82$-$2&  2.45$-$1& &4.16$-$2 & 2.26$-$1\\
m & 19& 3& 8.59$-$1&  5.01$+$0& &  8.67$-$1&  4.90$+$0& &  8.50$-$1&  5.46$+$0& &8.55$-$1 & 4.64$+$0\\
\hline
\end{tabular}
\end{table*}

\begin{table*}[p]
\centering
\caption[]{Effective collision strengths at $3.0\times 10^5$ K for transitions
from the ${\rm 1s}^2{\rm 2s}~^2{\rm S}_{1/2}$
ground level to the K-vacancy levels of {\rm Fe~{\sc xxiv}} showing the
effects of radiation and Auger dampings. ND: computed without damping.
RD: radiation damping is included. R+AD: radiation and Auger dampings are
included. Note: $a\pm b\equiv a\times 10^{\pm b}$.}
\label{table9}
\begin{tabular}{rrlll}
\hline\hline
$i$& $k$ & ND & RD & R+AD\\ \hline
1& 4 &2.96$-3$& 1.19$-3$ & 1.11$-3$ \\
1& 5 &1.26$-3$& 5.43$-4$ & 5.05$-4$ \\
1& 6 &2.23$-3$& 1.55$-3$ & 1.19$-3$ \\
1& 9 &3.19$-3$& 2.94$-3$ & 2.92$-3$ \\
1&10 &3.28$-5$& 1.60$-5$ & 1.49$-5$ \\
1&13 &6.36$-5$& 4.37$-6$ & 2.05$-6$ \\
1&14 &1.70$-5$& 6.41$-6$ & 2.61$-6$ \\
1&15 &4.07$-6$& 3.71$-6$ & 1.46$-6$ \\
1&17 &6.54$-6$& 6.11$-6$ & 1.98$-6$ \\
1&18 &3.34$-6$& 3.28$-6$ & 2.08$-6$ \\
\hline
\end{tabular}
\end{table*}

\begin{table*}[p]
\centering
\caption[]{Electron impact effective collision strengths  for transitions
within the $n=2$ complex of Fe~{\sc xxiv}}
\label{table10}
\begin{tabular}{rrllllllll}
\hline\hline
   &   & \multicolumn{7}{c}{Electron Temperature (K) } \\
\cline{3-10}\\
$i$& $k$ &1.00$+$5&5.00$+$5&1.00$+$6&5.00$+$6&1.00$+$7&5.00$+$7&1.00$+$8&$\infty$\\
\hline
  1&  4&1.13$-$3&1.09$-$3&1.06$-$3&1.03$-$3&1.03$-$3&1.04$-$3&1.06$-$3&1.16$-$3\\
  1&  5&5.18$-$4&4.98$-$4&4.88$-$4&4.52$-$4&4.19$-$4&2.89$-$4&2.23$-$4&3.50$-$5\\
  1&  6&1.31$-$3&1.13$-$3&1.06$-$3&9.62$-$4&9.01$-$4&6.89$-$4&6.02$-$4&2.41$-$4\\
  1&  7&1.42$-$3&1.40$-$3&1.39$-$3&1.31$-$3&1.21$-$3&7.90$-$4&5.69$-$4&0.0\\
  1&  8&1.08$-$3&1.08$-$3&1.09$-$3&1.16$-$3&1.23$-$3&1.82$-$3&2.24$-$3&2.44$-$3\\
  1&  9&2.92$-$3&2.94$-$3&2.99$-$3&3.26$-$3&3.55$-$3&5.66$-$3&7.25$-$3&7.93$-$3\\
  1& 10&1.51$-$5&1.48$-$5&1.47$-$5&1.47$-$5&1.46$-$5&1.45$-$5&1.46$-$5&1.52$-$5 \\
  1& 11&8.87$-$4&8.89$-$4&8.96$-$4&9.37$-$4&9.82$-$4&1.39$-$3&1.86$-$3&1.68$-$3\\
  1& 12&9.53$-$4&9.39$-$4&9.32$-$4&8.84$-$4&8.25$-$4&6.49$-$4&8.00$-$4&1.49$-$4\\
  1& 13&2.68$-$6&1.84$-$6&1.61$-$6&1.18$-$6&9.48$-$7&4.11$-$7&2.50$-$7&9.89$-$10 \\
  1& 14&2.79$-$6&2.44$-$6&2.21$-$6&1.68$-$6&1.38$-$6&6.64$-$7&4.46$-$7&2.24$-$8  \\
  1& 15&1.54$-$6&1.41$-$6&1.35$-$6&1.20$-$6&1.10$-$6&8.87$-$7&8.03$-$7&8.89$-$8  \\
  1& 16&7.42$-$6&7.30$-$6&7.29$-$6&7.24$-$6&7.13$-$6&6.93$-$6&6.89$-$6&6.49$-$6  \\
  1& 17&2.24$-$6&1.87$-$6&1.75$-$6&1.50$-$6&1.36$-$6&1.06$-$6&9.59$-$7&1.47$-$7  \\
  1& 18&2.14$-$6&2.05$-$6&2.01$-$6&1.81$-$6&1.64$-$6&1.25$-$6&1.09$-$6&2.32$-$8  \\
  1& 19&6.61$-$5&6.59$-$5&6.64$-$5&6.69$-$5&6.65$-$5&6.64$-$5&6.69$-$5&7.02$-$5  \\
  2&  4&1.05$-$4&9.45$-$5&8.49$-$5&7.19$-$5&7.04$-$5&8.04$-$5&9.05$-$5&8.66$-$5\\
  2&  5&2.98$-$4&2.90$-$4&2.73$-$4&2.38$-$4&2.21$-$4&1.57$-$4&1.20$-$4&9.35$-$7\\
  2&  6&6.54$-$4&5.56$-$4&4.91$-$4&3.86$-$4&3.51$-$4&2.45$-$4&1.86$-$4&1.94$-$13\\
  2&  7&7.35$-$5&5.72$-$5&3.92$-$5&1.13$-$5&6.22$-$6&1.52$-$6&8.14$-$7&2.51$-$14\\
  2&  8&1.26$-$3&1.25$-$3&1.25$-$3&1.30$-$3&1.35$-$3&1.56$-$3&1.71$-$3&2.24$-$3\\
  2&  9&1.55$-$4&1.18$-$4&1.03$-$4&8.46$-$5&7.78$-$5&5.50$-$5&4.22$-$5&2.42$-$13\\
  2& 10&7.59$-$4&7.49$-$4&7.42$-$4&7.07$-$4&6.66$-$4&5.06$-$4&4.36$-$4&1.64$-$4\\
  2& 11&1.78$-$4&1.74$-$4&1.74$-$4&1.78$-$4&1.82$-$4&2.05$-$4&2.20$-$4&2.82$-$4\\
  2& 12&1.16$-$4&7.73$-$5&6.79$-$5&5.70$-$5&5.27$-$5&3.72$-$5&2.83$-$5&1.62$-$13\\
  2& 13&9.15$-$4&8.93$-$4&8.83$-$4&8.31$-$4&7.69$-$4&5.02$-$4&3.63$-$4&1.60$-$6\\
  2& 14&1.04$-$3&9.85$-$4&9.67$-$4&9.04$-$4&8.36$-$4&5.46$-$4&3.94$-$4&6.32$-$12\\
  2& 15&2.22$-$3&2.23$-$3&2.27$-$3&2.43$-$3&2.60$-$3&3.97$-$3&5.18$-$3&5.31$-$3\\
  2& 16&1.68$-$3&1.69$-$3&1.72$-$3&1.87$-$3&2.04$-$3&3.27$-$3&4.33$-$3&4.56$-$3\\
  2& 17&4.28$-$4&4.22$-$4&4.20$-$4&3.96$-$4&3.67$-$4&2.40$-$4&1.73$-$4&5.38$-$11\\
  2& 18&1.23$-$4&1.23$-$4&1.24$-$4&1.28$-$4&1.32$-$4&1.71$-$4&2.09$-$4&1.92$-$4\\
  2& 19&1.07$-$4&1.06$-$4&1.07$-$4&1.05$-$4&1.02$-$4&9.87$-$5&1.05$-$4&8.14$-$5\\
  3&  4&2.74$-$4&2.20$-$4&1.66$-$4&8.81$-$5&7.60$-$5&7.91$-$5&8.95$-$5&8.80$-$5\\
  3&  5&7.84$-$5&7.02$-$5&6.01$-$5&3.01$-$5&2.22$-$5&1.16$-$5&8.42$-$6&3.33$-$14\\
  3&  6&4.08$-$4&3.06$-$4&2.52$-$4&1.65$-$4&1.43$-$4&9.54$-$5&7.29$-$5&4.03$-$6\\
  3&  7&1.06$-$3&1.05$-$3&9.54$-$4&7.80$-$4&7.15$-$4&5.01$-$4&3.82$-$4&8.88$-$14\\
  3&  8&8.97$-$5&6.41$-$5&5.03$-$5&3.28$-$5&2.88$-$5&1.95$-$5&1.50$-$5&6.36$-$14\\
  3&  9&1.58$-$3&1.45$-$3&1.38$-$3&1.34$-$3&1.34$-$3&1.45$-$3&1.53$-$3&1.98$-$3\\
  3& 10&2.38$-$4&2.24$-$4&2.17$-$4&1.99$-$4&1.84$-$4&1.21$-$4&8.78$-$5&1.53$-$6\\
  3& 11&2.87$-$4&2.56$-$4&2.37$-$4&2.10$-$4&1.95$-$4&1.39$-$4&1.06$-$4&4.63$-$13\\
  3& 12&2.42$-$3&2.09$-$3&2.01$-$3&2.00$-$3&2.04$-$3&2.31$-$3&2.49$-$3&3.07$-$3\\
  3& 13&1.29$-$3&1.11$-$3&1.06$-$3&9.65$-$4&8.97$-$4&6.42$-$4&5.22$-$4&1.38$-$4\\
  3& 14&2.54$-$3&2.21$-$3&2.12$-$3&1.96$-$3&1.86$-$3&1.56$-$3&1.48$-$3&8.04$-$4\\
  3& 15&1.12$-$3&1.10$-$3&1.09$-$3&1.06$-$3&1.01$-$3&8.96$-$4&8.82$-$4&5.78$-$4\\
  3& 16&7.07$-$4&7.03$-$4&7.09$-$4&7.41$-$4&7.75$-$4&1.07$-$3&1.35$-$3&1.34$-$3\\
  3& 17&3.72$-$3&3.70$-$3&3.72$-$3&3.80$-$3&3.88$-$3&4.83$-$3&5.83$-$3&5.33$-$3\\
  3& 18&4.37$-$3&4.40$-$3&4.47$-$3&4.78$-$3&5.12$-$3&7.79$-$3&1.01$-$2&1.04$-$2\\
  3& 19&1.29$-$3&1.29$-$3&1.30$-$3&1.34$-$3&1.37$-$3&1.78$-$3&2.18$-$3&2.07$-$3\\
\hline
\end{tabular}
\end{table*}
\end{singlespace}

\end{document}